%BEGIN MACRO
%%%
% This macro has been modified here and there by CGT.
% Last modification, 6/23/99.

%%                              JNL.TEX
%%
%%                This is JNL.TEX Version 0.3 as of 6/12/85.
%%
%%      This is a set of TeX 82 macros designed to produce scientific
%%      papers with a minimum of fuss and using as much of plain.tex as
%%      possible.  The user need only know what is in the TeXbook, and
%%      the macros under ``user definitions'' below.  Also, the user
%%      definitions are intended to be as simple as possible, so that
%%      the user may change them as desired.

%%
%%  Font definitions suitable for the IMAGEN (Written by Tony 
%%  Kennedy)
%%

%  Define a whole menagerie of pseudo-12pt fonts

\font\bigbf=cmbx10 scaled\magstep2

\font\twelverm=cmr10 scaled 1200    \font\twelvei=cmmi10 scaled 1200
\font\twelvesy=cmsy10 scaled 1200   \font\twelveex=cmex10 scaled 1200
\font\twelvebf=cmbx10 scaled 1200   \font\twelvesl=cmsl10 scaled 1200
\font\twelvett=cmtt10 scaled 1200   \font\twelveit=cmti10 scaled 1200

\skewchar\twelvei='177   \skewchar\twelvesy='60

%  Define \...point macros to change fonts and spacings consistently

\def\twelvepoint{\normalbaselineskip=12.4pt
  \abovedisplayskip 12.4pt plus 3pt minus 9pt
  \belowdisplayskip 12.4pt plus 3pt minus 9pt
  \abovedisplayshortskip 0pt plus 3pt
  \belowdisplayshortskip 7.2pt plus 3pt minus 4pt
  \smallskipamount=3.6pt plus1.2pt minus1.2pt
  \medskipamount=7.2pt plus2.4pt minus2.4pt
  \bigskipamount=14.4pt plus4.8pt minus4.8pt
  \def\rm{\fam0\twelverm}          \def\it{\fam\itfam\twelveit}%
  \def\sl{\fam\slfam\twelvesl}     \def\bf{\fam\bffam\twelvebf}%
  \def\mit{\fam 1}                 \def\cal{\fam 2}%
  \def\tt{\twelvett}
  \textfont0=\twelverm   \scriptfont0=\tenrm   \scriptscriptfont0=\sevenrm
  \textfont1=\twelvei    \scriptfont1=\teni    \scriptscriptfont1=\seveni
  \textfont2=\twelvesy   \scriptfont2=\tensy   \scriptscriptfont2=\sevensy
  \textfont3=\twelveex   \scriptfont3=\twelveex 
 \scriptscriptfont3=\twelveex
  \textfont\itfam=\twelveit
  \textfont\slfam=\twelvesl
  \textfont\bffam=\twelvebf \scriptfont\bffam=\tenbf
  \scriptscriptfont\bffam=\sevenbf
  \normalbaselines\rm}

%       tenpoint

%%
%%      Various internal macros
%%

\def\beginlinemode{\endmode
  \begingroup\parskip=0pt 
\obeylines\def\\{\par}\def\endmode{\par\endgroup}}
\def\beginparmode{\endmode
  \begingroup \def\endmode{\par\endgroup}}
\let\endmode=\par
{\obeylines\gdef\
{}}
\def\singlespace{\baselineskip=\normalbaselineskip}
\def\oneandathirdspace{\baselineskip=\normalbaselineskip
  \multiply\baselineskip by 4 \divide\baselineskip by 3}

\def\doublespace{\baselineskip=
\normalbaselineskip \multiply\baselineskip by 2}

\newcount\firstpageno
\firstpageno=1
\footline={\ifnum\pageno<\firstpageno{\hfil}%
\else{\hfil\twelverm\folio\hfil}\fi}
\let\rawfootnote=\footnote              % We must set the footnote style
\def\footnote#1#2{{\rm\singlespace\parindent=0pt\rawfootnote{#1}{#2}}}
\def\raggedcenter{\leftskip=4em plus 12em \rightskip=\leftskip
  \parindent=0pt \parfillskip=0pt \spaceskip=.3333em \xspaceskip=.5em
  \pretolerance=9999 \tolerance=9999
  \hyphenpenalty=9999 \exhyphenpenalty=9999 }
\def\dateline{\rightline{\ifcase\month\or
  January\or February\or March\or April\or May\or June\or
  July\or August\or September\or October\or November\or December\fi
  \space\number\year}}
\def\received{\vskip 3pt plus 0.2fill
 \centerline{\sl (Received\space\ifcase\month\or
  January\or February\or March\or April\or May\or June\or
  July\or August\or September\or October\or November\or December\fi
  \qquad, \number\year)}}

%%
%%      Page layout, margins, font and spacing (feel free to change)
%%

\hsize=6.5truein
%\hoffset=1truein
\vsize=8.9truein
\voffset=0.0truein
\parskip=\medskipamount
\twelvepoint            % selects twelvepoint fonts (cf. \tenpoint)
\oneandathirdspace           % selects spacing for main part of paper 
                        %      (cf. \singlespace, \oneandahalfspace)
\overfullrule=0pt       % delete the nasty little black boxes for overfull box

%%
%%      The user definitions for major parts of a paper 
%%		(feel free to change)
%%

    % Preprint number at upper right of title page

\def\title                      %  Title on title page
  {\null\vskip 3pt plus 0.2fill
   \beginlinemode \doublespace \raggedcenter \bigbf}

\def\author                     %  Author(s) name(s)  on title page
  {\vskip 3pt plus 0.2fill \beginlinemode
   \singlespace \raggedcenter}

\def\affil                      % Affiliations (can intermix with \author)
  {\vskip 4pt %plus 0.1fill 
\beginlinemode
   \singlespace \raggedcenter \sl}

\def\abstract                   % Begin abstract
  {\vskip 3pt plus 0.3fill \beginparmode
   \oneandathirdspace\narrower}

\def\endtitlepage               % End title page, begin body of paper
  {\endpage                     %       This subsumes \body
   \body}

\def\body                       % Begin text body;  can be used to end
  {\beginparmode}               % \title, \author, \affil, \abstract,
                                % \reference, or \figurecaption modes

\def\head#1{                    % Head;  NOTE enclose the text in {}
  \vskip 0.25truein     %  e.g., \head{I. Introduction}
 {\immediate\write16{#1}
   \noindent{\bf{#1}}\par}
   \nobreak\vskip 0.125truein\nobreak}

\def\subhead#1{                 % Subhead;  NOTE enclose the text in {}
  \vskip 0.25truein             % e.g., \subhead{A. History of the Problem}
  \noindent{{\it {#1}} \par}
   \nobreak\vskip 0.15truein\nobreak}

\def\refto#1{[#1]}           % For references in text using square brackets

\def\references                 % Begin references -- basic format is Phys Rev
  {\subhead{\bf References}         % I.e., volume, page, year 
%					(space after commas)
   \beginparmode
   \frenchspacing \parindent=0pt \leftskip=1truecm
   \oneandathirdspace\parskip=8pt plus 3pt
 \everypar{\hangindent=\parindent}}

\gdef\refis#1{\indent\hbox to 0pt{\hss#1.~}}    % Ref list numbers.

\gdef\journal#1, #2, #3, #4#5#6#7{               % Journal reference.  
%%								Comma set
    {\sl #1~}{\bf #2}, #3 (#4#5#6#7)}           % off: name, vol, page, year

\def\refstylenp{                % Nucl Phys(or Phys Lett) ref style: V, Y, P
  \gdef\refto##1{ [##1]}                                % Reference in text []
  \gdef\refis##1{\indent\hbox to 0pt{\hss##1)~}}        % Ref list numbers)
  \gdef\journal##1, ##2, ##3, ##4 {                     % Journal reference
     {\sl ##1~}{\bf ##2~}(##3) ##4 }}

\def\refstyleprnp{              % Input like pr, output like np!!
  \gdef\refto##1{ [##1]}                                % Reference in text []
  \gdef\refis##1{\indent\hbox to 0pt{\hss##1)~}}        % Ref list numbers)
  \gdef\journal##1, ##2, ##3, 1##4##5##6{               % Journal reference
    {\sl ##1~}{\bf ##2~}(1##4##5##6) ##3}}

\def\prd{\journal Phys. Rev. D, }

\def\jmp{\journal J. Math. Phys., }

\def\cmp{\journal Commun. Math. Phys., }

\def\cqg{\journal Class. Quantum Grav., }

\def\endreferences{\body}

\def\figurecaptions             % Begin figure captions
  { \beginparmode
   \subhead{Figure Captions}
}

\def\endpage                    %  Eject a page
  {\vfill\eject}

\def\endpaper                   %  Ways to say goodbye
  {\endmode\vfill\supereject}

%%
%%      Various little user definitions
%%
%\hook is the inner evaluation macro
\def\hook{\mathbin{\raise2.5pt\hbox{\hbox{{\vbox{\hrule height.4pt 
width6pt depth0pt}}}\vrule height3pt width.4pt depth0pt}\,}}
\def\today{\number\day\ \ifcase\month\or
     January\or February\or March\or April\or May\or June\or
     July\or August\or September\or October\or November\or
     December\space \fi\ \number\year}
\def\date{\noindent{\tt 
     Date typeset: \today\par\bigskip}}
\def\ref#1{Ref. #1}                     %       for inline references
\def\Ref#1{Ref. #1}                     %       ditto

          % For citation of equation numbers
        %       ditto
                     %       ditto
                     %       ditto
                   %       ditto
                   %       ditto
\def\frac#1#2{{\textstyle{#1 \over #2}}}
\def\half{{\textstyle{ 1\over 2}}}
\def\>{\rangle}
\def\<{\langle}
\def\eg{{\it e.g.,\ }}

\def\ie{{\it i.e.,\ }}

\def\etc{{\it etc.}}

\def\sla{\raise.15ex\hbox{$/$}\kern-.57em}
\def\leaderfill{\leaders\hbox to 1em{\hss.\hss}\hfill}
\def\twiddle{\lower.9ex\rlap{$\kern-.1em\scriptstyle\sim$}}
\def\bigtwiddle{\lower1.ex\rlap{$\sim$}}
\def\gtwid{
\mathrel{\raise.3ex\hbox{$>$\kern-.75em\lower1ex\hbox{$\sim$}}}}
\def\ltwid{\mathrel{\raise.3ex\hbox
{$<$\kern-.75em\lower1ex\hbox{$\sim$}}}}
\def\square{\kern1pt\vbox{\hrule height 1.2pt\hbox
{\vrule width 1.2pt\hskip 3pt
   \vbox{\vskip 6pt}\hskip 3pt\vrule width 0.6pt}
\hrule height 0.6pt}\kern1pt}

\def\m@th{\mathsurround=0pt }
\def\leftrightarrowfill{$\m@th \mathord\leftarrow \mkern-6mu
 \cleaders\hbox{$\mkern-2mu \mathord- \mkern-2mu$}\hfill
 \mkern-6mu \mathord\rightarrow$}
\def\overleftrightarrow#1{\vbox{\ialign{##\crcr
     \leftrightarrowfill\crcr\noalign{\kern-1pt\nointerlineskip}
     $\hfil\displaystyle{#1}\hfil$\crcr}}}

%% *********** New stuff follows *******************

\font\titlefont=cmr10 scaled\magstep3

\def\martinstyletitle                      %  Title on title page
  {\null\vskip 3pt plus 0.2fill
   \beginlinemode \doublespace \raggedcenter \titlefont}

\font\twelvesc=cmcsc10 scaled 1200

\def\author                     %  Author(s) name(s)  on title page
  {\vskip 3pt plus 0.2fill \beginlinemode
   \singlespace \raggedcenter\twelvesc}

%%
%%      AmSTeX compatability definitions
%%
%%      To run a TeX file originally intended for AmSTeX, 
%%      only small changes
%%      should be necessary (I hope).  Use the line \input jnl at the start.
%%      Remove the lines \input amstex, \documentstyle{itpjnl} at the
%%      beginning;  also remove all the page layout stuff (\parindent=1cm,
%%      \hsize=5.28125in etc.)  The page layout is now done automatically.
%%      Also OMIT the qualifier \magnification=1200 when you 
%%      IMPRINT the
%%      .dvi file.  (\TagsOnRight is harmless, you can take it out or leave
%%      it in.)  I believe most AmSTeX will work with no change.  
%%      One problem
%%      is \footnote, which is a little different in that it now needs to
%%      have an explicit asterisk *  (or whatever) included, like this:
%%              \footnote*{Text winds up at bottom of page.}
%%      This is discussed on p. 116 of the TeXbook.  
%%      IGNORE the AmSTeX
%%      documentation (if you can call it that);  refer to the TeXbook.
%%
%%      Note that many commands in AmSTeX have their equivalents 
%%      in the
%%      TeXbook, perhaps with different names and slightly differing
%%      usage. E.g., the old \align in AmSTeX is replaced by \eqalign
%%      (p. 190) and \aligntag is replaced by \eqalignno (p. 192).
%%      \align and \aligntag still work, but I recommend that you use
%%      \eqalign and \eqalignno in documents run under jnl.
%%
%%      See me if you have any problems  -- Doug.
%%

\def\endtitle{\body}

\def\heading                            % Heading
  {\vskip 0.5truein plus 0.1truein      % e.g., \heading I. NOTES 
\endheading
   \beginparmode \def\\{\par} \parskip=0pt \singlespace \raggedcenter}

\def\endheading
  {\par\nobreak\vskip 0.25truein\nobreak\beginparmode}

\def\subheading                         % Subheading
  {\vskip 0.25truein plus 0.1truein     
% e.g., \subheading{A. The Problem}
   \beginlinemode \singlespace \parskip=0pt \def\\{\par}\raggedcenter}

\def\tag#1$${\eqno(#1)$$}

\def\align#1$${\eqalign{#1}$$}

\def\aligntag#1$${\gdef\tag##1\\{&(##1)\cr}\eqalignno{#1\\}$$
  \gdef\tag##1$${\eqno(##1)$$}}

\def\endaligntag{}

\def\overset #1\to#2{{\mathop{#2}\limits^{#1}}}
\def\underset#1\to#2{{\let\next=#1\mathpalette\undersetpalette#2}}
\def\undersetpalette#1#2{\vtop{\baselineskip0pt
\ialign{$\mathsurround=0pt #1\hfil##\hfil$\crcr#2\crcr\next\crcr}}}

%%
%%      Various little user definitions
%%

\def\ref#1{Ref.~#1}                     %       for inline references
\def\Ref#1{Ref.~#1}                     %       ditto
\def\[#1]{[\cite{#1}]}
\def\cite#1{{#1}}
%%\def\Equation#1{Equation~(#1)}                % For citation of 
%equation numb
%%\def\Equations#1{Equations~(#1)}      %       ditto
%%\def\Eq#1{Eq.~(#1)}                   %       ditto
%%\def\Eqs#1{Eqs.~(#1)}                 %       ditto
\def\(#1){(\call{#1})}
\def\call#1{{#1}}
\def\taghead#1{}
\def\frac#1#2{{#1 \over #2}}
\def\half{{\frac 12}}

\def\12{{1\over2}}
\def\eg{{\it e.g.,\ }}

\def\ie{{\it i.e.,\ }}

\def\etc{{\it etc.\ }}

\def\cf{{\sl cf.\ }}
\def\sla{\raise.15ex\hbox{$/$}\kern-.57em}
\def\leaderfill{\leaders\hbox to 1em{\hss.\hss}\hfill}
\def\twiddle{\lower.9ex\rlap{$\kern-.1em\scriptstyle\sim$}}
\def\bigtwiddle{\lower1.ex\rlap{$\sim$}}
\def\gtwid{\mathrel{\raise.3ex\hbox{$>$
\kern-.75em\lower1ex\hbox{$\sim$}}}}
\def\ltwid{\mathrel{\raise.3ex\hbox{$<$
\kern-.75em\lower1ex\hbox{$\sim$}}}}
\def\square{\kern1pt\vbox{\hrule height 1.2pt\hbox
{\vrule width 1.2pt\hskip 3pt
   \vbox{\vskip 6pt}\hskip 3pt\vrule width 0.6pt}
\hrule height 0.6pt}\kern1pt}
\def\tdot#1{\mathord{\mathop{#1}\limits^{\kern2pt\ldots}}}

\def\pmb#1{\setbox0=\hbox{#1}%
  \kern-.025em\copy0\kern-\wd0
  \kern  .05em\copy0\kern-\wd0
  \kern-.025em\raise.0433em\box0 }

\catcode`@=11
\newcount\tagnumber\tagnumber=0

\immediate\newwrite\eqnfile
\newif\if@qnfile\@qnfilefalse
\def\write@qn#1{}
\def\writenew@qn#1{}
\def\w@rnwrite#1{\write@qn{#1}\message{#1}}
\def\@rrwrite#1{\write@qn{#1}\errmessage{#1}}

\def\taghead#1{\gdef\t@ghead{#1}\global\tagnumber=0}
\def\t@ghead{}

\expandafter\def\csname @qnnum-3\endcsname
  {{\t@ghead\advance\tagnumber by -3\relax\number\tagnumber}}
\expandafter\def\csname @qnnum-2\endcsname
  {{\t@ghead\advance\tagnumber by -2\relax\number\tagnumber}}
\expandafter\def\csname @qnnum-1\endcsname
  {{\t@ghead\advance\tagnumber by -1\relax\number\tagnumber}}
\expandafter\def\csname @qnnum0\endcsname
  {\t@ghead\number\tagnumber}
\expandafter\def\csname @qnnum+1\endcsname
  {{\t@ghead\advance\tagnumber by 1\relax\number\tagnumber}}
\expandafter\def\csname @qnnum+2\endcsname
  {{\t@ghead\advance\tagnumber by 2\relax\number\tagnumber}}
\expandafter\def\csname @qnnum+3\endcsname
  {{\t@ghead\advance\tagnumber by 3\relax\number\tagnumber}}

\def\equationfile{%
  \@qnfiletrue\immediate\openout\eqnfile=\jobname.eqn%
  \def\write@qn##1{\if@qnfile\immediate\write\eqnfile{##1}\fi}
  \def\writenew@qn##1{\if@qnfile\immediate\write\eqnfile
    {\noexpand\tag{##1} = (\t@ghead\number\tagnumber)}\fi}
}

\def\callall#1{\xdef#1##1{#1{\noexpand\call{##1}}}}
\def\call#1{\each@rg\callr@nge{#1}}

\def\each@rg#1#2{{\let\thecsname=#1\expandafter\first@rg#2,\end,}}
\def\first@rg#1,{\thecsname{#1}\apply@rg}
\def\apply@rg#1,{\ifx\end#1\let\next=\relax%
\else,\thecsname{#1}\let\next=\apply@rg\fi\next}

\def\callr@nge#1{\calldor@nge#1-\end-}
\def\callr@ngeat#1\end-{#1}
\def\calldor@nge#1-#2-{\ifx\end#2\@qneatspace#1 %
  \else\calll@@p{#1}{#2}\callr@ngeat\fi}
\def\calll@@p#1#2{\ifnum#1>#2{\@rrwrite
{Equation range #1-#2\space is bad.}
\errhelp{If you call a series of equations by the notation M-N, then M and
N must be integers, and N must be greater than or equal to M.}}\else %
{\count0=#1\count1=
#2\advance\count1 by1\relax\expandafter\@qncall\the\count0,%
  \loop\advance\count0 by1\relax%
    \ifnum\count0<\count1,\expandafter\@qncall\the\count0,%
  \repeat}\fi}

\def\@qneatspace#1#2 {\@qncall#1#2,}
\def\@qncall#1,{\ifunc@lled{#1}{\def\next{#1}\ifx\next\empty\else
  \w@rnwrite{Equation number \noexpand\(>>#1<<) 
has not been defined yet.}
  >>#1<<\fi}\else\csname @qnnum#1\endcsname\fi}

\let\eqnono=\eqno
\def\eqno(#1){\tag#1}
\def\tag#1$${\eqnono(\displayt@g#1 )$$}

\def\aligntag#1\endaligntag
  $${\gdef\tag##1\\{&(##1 )\cr}\eqalignno{#1\\}$$
  \gdef\tag##1$${\eqnono(\displayt@g##1 )$$}}

\def\eqalignno#1{\displ@y \tabskip\centering
  \halign to\displaywidth{\hfil$\displaystyle{##}$\tabskip\z@skip
    &$\displaystyle{{}##}$\hfil\tabskip\centering
    &\llap{$\displayt@gpar##$}\tabskip\z@skip\crcr
    #1\crcr}}

\def\displayt@gpar(#1){(\displayt@g#1 )}

\def\displayt@g#1 {\rm\ifunc@lled{#1}\global\advance\tagnumber by1
        {\def\next{#1}\ifx\next\empty\else\expandafter
        \xdef\csname
 @qnnum#1\endcsname{\t@ghead\number\tagnumber}\fi}%
  \writenew@qn{#1}\t@ghead\number\tagnumber\else
        {\edef\next{\t@ghead\number\tagnumber}%
        \expandafter\ifx\csname @qnnum#1\endcsname\next\else
        \w@rnwrite{Equation \noexpand\tag{#1} is 
a duplicate number.}\fi}%
  \csname @qnnum#1\endcsname\fi}

\def\ifunc@lled#1{\expandafter\ifx\csname @qnnum#1\endcsname\relax}

\let\@qnend=\end\gdef\end{\if@qnfile
\immediate\write16{Equation numbers 
written on []\jobname.EQN.}\fi\@qnend}

\catcode`@=12

\catcode`@=11
\newcount\r@fcount \r@fcount=0
\newcount\r@fcurr
\immediate\newwrite\reffile
\newif\ifr@ffile\r@ffilefalse
\def\w@rnwrite#1{\ifr@ffile\immediate\write\reffile{#1}\fi\message{#1}}

\def\writer@f#1>>{}
\def\referencefile{%			  Stuff to write .REF file
  \r@ffiletrue\immediate\openout\reffile=\jobname.ref%
  \def\writer@f##1>>{\ifr@ffile\immediate\write\reffile%
    {\noexpand\refis{##1} = \csname r@fnum##1\endcsname = %
     \expandafter\expandafter\expandafter\strip@t\expandafter%
     \meaning\csname r@ftext
\csname r@fnum##1\endcsname\endcsname}\fi}%
  \def\strip@t##1>>{}}

\def\citeall#1{\xdef#1##1{#1{\noexpand\cite{##1}}}}
\def\cite#1{\each@rg\citer@nge{#1}}	% Variable No. of args,
% separated by

\def\each@rg#1#2{{\let\thecsname=#1\expandafter\first@rg#2,\end,}}
\def\first@rg#1,{\thecsname{#1}\apply@rg}	% each@ag is a general
% purpose
\def\apply@rg#1,{\ifx\end#1\let\next=\relax%	  variable no. of arg.
 %macro.
\else,\thecsname{#1}\let\next=\apply@rg\fi\next}% args separated 
%by commas

\def\citer@nge#1{\citedor@nge#1-\end-}	% Check for M-N range 
%(M and N numbers)
\def\citer@ngeat#1\end-{#1}
\def\citedor@nge#1-#2-{\ifx\end#2\r@featspace#1 % Single argument
  \else\citel@@p{#1}{#2}\citer@ngeat\fi}	% M-N range of arguments
\def\citel@@p#1#2{\ifnum#1>#2{\errmessage{Reference range #1-
#2\space is bad.}%
    \errhelp{If you cite a series of references by the notation M-N, then M 
and
    N must be integers, and N must be greater than or equal to M.}}\else%
 {\count0=#1\count1=#2\advance\count1 
by1\relax\expandafter\r@fcite\the\count0,
  \loop\advance\count0 by1\relax%	  Loop from M to N
    \ifnum\count0<\count1,\expandafter\r@fcite\the\count0,%
  \repeat}\fi}

\def\r@featspace#1#2 {\r@fcite#1#2,}	% Eat spaces at beginning or 
%end of arg
\def\r@fcite#1,{\ifuncit@d{#1}%		  Cite individual reference
    \newr@f{#1}%
    \expandafter\gdef\csname r@ftext\number\r@fcount\endcsname%
                     {\message{Reference #1 to be supplied.}%
                      \writer@f#1>>#1 to be supplied.\par}%
 \fi%
 \csname r@fnum#1\endcsname}
\def\ifuncit@d#1{\expandafter\ifx\csname r@fnum#1\endcsname\relax}%
\def\newr@f#1{\global\advance\r@fcount by1%
    \expandafter\xdef\csname r@fnum#1\endcsname{\number\r@fcount}}

\let\r@fis=\refis			% Save old \refis, redefine
\def\refis#1#2#3\par{\ifuncit@d{#1}%      Use two params #2 #3 to 
%strip blank
   \newr@f{#1}%
   \w@rnwrite{Reference #1=\number\r@fcount\space is not cited up to
 now.}\fi%
  \expandafter
\gdef\csname r@ftext\csname r@fnum#1\endcsname\endcsname%
  {\writer@f#1>>#2#3\par}}

\def\ignoreuncited{%   redefine \refis if ignoring uncited references
   \def\refis##1##2##3\par{\ifuncit@d{##1}%
    \else\expandafter\gdef
\csname r@ftext\csname r@fnum##1\endcsname\endcsname%
     {\writer@f##1>>##2##3\par}\fi}}

\def\r@ferr{\endreferences\errmessage{I was expecting to see
\noexpand\endreferences before now;  I have inserted it here.}}
\let\r@ferences=\references
\def\references{\r@ferences\def\endmode{\r@ferr\par\endgroup}}

\let\endr@ferences=\endreferences
\def\endreferences{\r@fcurr=0%		  Save old \endreferences, 
%redefine
  {\loop\ifnum\r@fcurr<\r@fcount%	  Loop over refnum and 
%produce text
    \advance\r@fcurr by 
1\relax\expandafter\r@fis\expandafter{\number\r@fcurr}%
    \csname r@ftext\number\r@fcurr\endcsname%
  \repeat}\gdef\r@ferr{}\endr@ferences}

% Save old \endpaper, redefine it to write parting message.

\let\r@fend=\endpaper\gdef\endpaper{\ifr@ffile
\immediate\write16{Cross References written on 
[]\jobname.REF.}\fi\r@fend}

\catcode`@=12

\citeall\refto		% These macros will generate citations
\citeall\ref		%
\citeall\Ref		%

% END MACRO

\ignoreuncited
\pageno=0

\def\q{\hat q}
\def\h{{\bf h}}
\def\n{{\bf n}}
\def\m{{\bf m}}
\def\G{{\cal G}}
\def\Q{{\cal Q}}
\def\Qhat{{\cal Q}^{\ss G}}

\def\ss{\scriptscriptstyle}
\def\s{{\cal S}}
\def\sg{{\cal S}^{\ss G}}
\def\sgperp{({\cal S}_{x}^{\ss G_{x}})^{\perp}}
\def\sgx{S_{x}^{\ss G_{x}}}
\def\sgstarx{(S^{*}_{x})^{\ss G_{x}}}
\def\sgxo{(S_{x}^{\ss G_{x}})^{0}}
\def\proof{\smallskip\noindent
{\bf Proof:\ }}
\def\reals{{{\rm I \!\!\! R}}}

\line{\hfill August 2001}

\title
THE PRINCIPLE OF SYMMETRIC CRITICALITY IN GENERAL RELATIVITY
\endtitle

\bigskip

\author
Mark E. Fels 
\affil
Department of Mathematics and Statistics

Utah State University, Logan, Utah 84322

\author
Charles G. Torre
\affil
Department of Physics

Utah State University, Logan, Utah 84322

\abstract
\noindent{\bf Abstract}

We consider  a version of Palais' Principle of Symmetric Criticality  (PSC) 
that is
applicable to the Lie symmetry reduction  of
Lagrangian field theories.  PSC asserts that, given a group action, 
for any group-invariant Lagrangian
the equations obtained by restriction of 
Euler-Lagrange equations to group-invariant fields 
 are equivalent to the Euler-Lagrange equations of a canonically defined, 
symmetry-reduced  Lagrangian. 
We investigate the validity of PSC 
for local gravitational theories built from a metric.  It is shown 
that there are two independent conditions which must be satisfied for 
PSC to be valid. 
One of these conditions, obtained previously in the 
context of transverse symmetry group actions, provides a generalization 
of the well-known 
unimodularity condition that arises in spatially homogeneous cosmological 
models. 
The 
other condition seems to be new.
The conditions that determine the validity of PSC are equivalent to pointwise 
conditions on the group action alone.  These results are illustrated 
with a variety of examples from general relativity. It is straightforward 
to generalize all
of our results to any relativistic field theory.

\endtitlepage

\vfill\eject
\taghead{1.}
\head{1. Introduction}

An important approach to studying properties of the solution space 
of gravitational field equations is to restrict attention to metrics and 
matter fields that admit a specified group of symmetries. It was noted 
quite some time ago 
by Hawking \refto{Hawking1969},
 and subsequently discussed in some detail by MacCallum and 
Taub \refto{MacCallum1972},
 that with homogeneous cosmological models (``Bianchi models'') 
one cannot always impose the symmetry on the fields in the 
Einstein-Hilbert action functional since varying the action in this 
restricted class of fields may not yield the correct field equations. 
In particular, it was noted that  only the Bianchi class A groups 
would, in general, allow for a successful symmetry reduction 
of the Einstein-Hilbert action. Many others have elaborated 
on this issue in homogeneous cosmology, see, for example, 
\refto{Ryan1974, Sneddon1976,  
MacCallum1979, Jantzen1980, Ashtekar1991c, Shepley1998} and 
references therein. As Hawking points out in \refto{Hawking1969}, 
the difficulty which arises with the Bianchi class B models is 
due to the presence of a non-trivial boundary term in the 
restricted variational principle. Such a difficulty does not 
appear in many other symmetry reductions.  For example,
Pauli restricts the Einstein-Hilbert action to a class of static, 
spherically symmetric metrics and obtains the reduced 
Einstein equations by Hamilton's principle \refto{Pauli1921} 
(he attributes this approach to Weyl).   In addition, Lovelock 
has shown that a variety of  Lagrangians for 
fourth-order field equations allow for reduction by spherical 
symmetry  \refto{Lovelock1973}.  In light of  such examples, 
it is natural to ask whether there exist general criteria that allow 
one to decide for a given symmetry group when one can successfully 
reduce a generic Lagrangian or action functional. Our goal in this 
paper is to give a systematic account of the 
symmetry reduction of gravitational Lagrangians and field equations, 
and to completely characterize the symmetry 
group actions that guarantee the reduced Lagrangian produces the 
reduced field equations.

Viewing these issues strictly from the 
point of view of action functionals and variational principles
Palais has arrived at  
 the {\it Principle of Symmetric 
Criticality} (PSC) \refto{Palais1979}.  
Given a group action on a space of fields,  
one can consider the restriction of an action functional $ S$ 
to the group invariant fields to obtain the reduced action $\hat S$.
Palais'  PSC asserts that, for any group invariant functional $S$, 
critical points of $\hat S$ within the class of group invariant fields 
are (group invariant) critical points of $S$.
As Palais emphasized, PSC need be neither well-defined nor valid. 
Under hypotheses  that guarantee PSC makes 
sense, he goes on to give necessary and sufficient conditions for the 
validity of PSC in a variety of settings. 
Unfortunately, a 
straightforward 
application of these results to general relativity (and, more 
generally, classical field theory) is somewhat awkward
since one must decide at the outset what class of spacetimes to 
consider in the variational principle, what asymptotic and/or 
boundary conditions to impose, what to do about spacetime 
singularities, \etc Moreover, different group actions may necessitate 
different choices in this regard. All these issues, which are 
fundamentally global in nature, will arise when using 
PSC formulated
 in terms of the action integral, viewed
as a functional on the infinite-dimensional space of 
metrics. 

These difficulties can be avoided by using a purely local
formulation of PSC that is based on the Lagrangian rather than on the 
action integral. 
The version of PSC adopted in this paper asserts that, for a given group action and for any group invariant Lagrangian, the reduced field equations obtained by restriction of Euler-Lagrange equations to group invariant fields  are equivalent to the 
Euler-Lagrange equations of a canonically defined, reduced Lagrangian.
This formulation of PSC 
was studied  in \refto{Anderson1997} under the  hypothesis of a  transverse symmetry group action. Using tools developed in \refto{CGT2000}, this formulation of PSC has been extended to the general, non-transverse case in \refto{AFT2000}.  As we shall see, by using a purely local 
formulation of PSC based upon the Lagrangian,
it is possible to give 
necessary and sufficient conditions for the validity of PSC once and 
for all, without having to wrestle with the complications mentioned in 
the previous paragraph.

In the context of an arbitrary metric theory of gravity in any number of 
dimensions, 
we shall show 
that there are two independent conditions which determine the validity of PSC. 
One of these 
 already appears for transverse group actions 
in \refto{Anderson1997}; it generalizes the restriction to Bianchi class 
A in the case  of
homogeneous cosmological models, which was noted above.  
The second 
condition for PSC is only relevant when
considering  non-transverse group actions  \refto{AFT2000} and does not 
seem to have been treated in the physics literature on symmetry reduction 
of variational principles. 
The conditions that determine the validity of PSC are equivalent to 
pointwise conditions on the group action. Thus, the validity of PSC is 
determined solely by the group action,
 irrespective of the Lagrangian, the spacetime manifold, or the choice of 
asymptotic or boundary conditions. It is straightforward to generalize 
these results to other settings, \eg a field theory that includes 
other fields besides the metric, or for a theory of matter fields on a 
fixed spacetime. 

We check the validity of PSC for a number of examples from general relativity. 
The group actions used in these examples have largely been taken 
from  \refto{Petrov1969}. There,  
Petrov has provided a (not quite complete) 
classification of 
Killing vector fields in four dimensions.  It is a simple exercise to check 
PSC  for any of the vector field systems appearing in \refto{Petrov1969} 
by following the pattern of the
examples
presented here.

It should be emphasized at the outset that the symmetry reductions 
we consider do not involve any ``coordinate conditions'' or ``gauge 
fixing conditions''.  In what follows we will always work with the most 
general metrics admitting the chosen isometry groups, and it is in 
this setting that we discuss the validity of PSC. Of course, if metric 
components are eliminated from the Lagrangian using coordinate 
conditions one will not, in general, recover the corresponding 
Euler-Lagrange equations.  It is interesting to note, however, 
that for certain symmetry reductions there exist privileged classes 
of coordinate conditions which {\it can} be imposed in the Lagrangian 
without causing a net loss of independent reduced field equations.  
See \refto{Pauli1921, Lovelock1973} for illustrations of this phenomenon.

This paper is organized as follows. 
Section 2 summarizes the prerequisites from the theory of isometry 
groups and the construction of group invariant metrics. 
Section 3 defines group-invariant Lagrangians and field equations. 
Section 4 details the
construction of the 
reduced equations and reduced Lagrangian. Section 5 gives a 
precise formulation of PSC and derives the necessary and sufficient 
conditions for its validity.  Section 6  provides a variety of 
examples that illustrate the results 
presented in the previous sections.
Appendix A provides a brief description of the generalization of our results to the 
case where other types of fields are included, either coupled dynamically to 
the metric or 
propagating on a fixed spacetime. Appendix B gives a 
precise notion of equivalence of differential equations which we use 
to formulate PSC.

\taghead{2.}
\head{2. Metrics with Symmetry}

Often in general relativity one fixes a spacetime $(M,g)$ and determines
its isometry group (or algebra), \eg by solving the Killing equations for the
Killing vector fields. However, when considering symmetry reduction 
of a gravitational theory we take the opposite point of view and are 
interested in 
restricting attention to all spacetimes which admit a {\it chosen} 
isometry group (or algebra).  In order to parameterize the spacetime 
metrics which admit a 
specified isometry group we identify a bundle over $M/G$ whose 
sections
are in one to one correspondence with the sought after spacetimes. 
In this section we describe the construction of this bundle. A short 
summary of the infinitesimal version of the problem is also given 
at the end of the section.  Further details can be found in  \refto{CGT2000}.

Let $\s$ be the space of smooth symmetric tensor fields 
of type $\left({}^{0}_{2}\right)$
on the manifold $M$, 
and denote by $\Q\subset \s$
the subset of smooth Lorentz metrics. Let
$$
\mu \, \colon G\times M\to M,
\tag gaction
$$
be an action of the Lie group $G$ on $M$. The group
$G$ acts naturally on $\s$ and $\Q$ by pull-back:
$$
g \to \mu_\gamma ^{*}g,\quad g\in \s,\quad \gamma \in G,
$$
where $\mu_\gamma :M \to M$ is the diffeomorphism obtained by 
restricting $\mu$ in \(gaction) to a fixed element $\gamma \in G$. 
A symmetric tensor $g \in \s $ is {\it $G$-invariant} if 
$$
\mu_{\gamma}^{*} g = g \quad \forall\ \gamma\in G. 
\tag metric_symmetry
$$
Let $\sg \subset \s$ be the subset of $G$-invariant symmetric 
$\left({}^{0}_{2}\right)$ tensor fields on $M$, and let $\Qhat \subset \sg$ 
the $G$-invariant metrics. Thus each $g \in \Qhat$ admits $G$ as an isometry group.

The passage from  $\Q$ to $\Qhat$ according to the $G$-invariance condition 
\(metric_symmetry)  involves 
two types of ``reduction''. First of all, there is 
 the familiar ``dimensional reduction'' in the number of 
independent variables from the spacetime dimension to the codimension 
of the orbits of the group in $M$. Thus the set of 
$G$-invariant metrics can be parametrized by fields on the {\it reduced 
spacetime} $M/G$. Henceforth we assume the quotient space $M/G$ is 
a smooth manifold.
The second type of reduction 
determines the number of independent components of the metric
(or fields) that are essential in parametrizing $\Qhat$.
The  procedure which determines the reduction of the metric components will
be used throughout the article so we present some details. 

Let $G$ be a Lie group acting on $M$, the 
{\it isotropy group} $G_x $ of a point $x\in M$ is the subgroup 
$$
G_{x} = \{ \ \gamma \in G \quad | \quad   \mu(\gamma,x) = x\  \}\ .
$$
If the isotropy group $G_{x}$ has dimension $p$ and the group itself 
has dimension $d$, then the orbit through $x$ has dimension $l=d-p$. 
The isotropy group $G_{x}$ acts on the tangent space $T_xM$ at $x$ 
by the push-forward map $\mu_{\gamma *} : T_xM \to T_xM$. That is, 
given $\gamma\in G_{x}$ and $V\in T_{x}M$, then
$$
(\gamma, V) \to \mu_{\gamma*} V. 
$$
The homomorphism $\gamma \to \mu_{\gamma *}$ of $G_{x} \to 
GL(T_xM)$ is the {\it linear
isotropy representation} of the group $G_{x}$. Consequently
there is the induced representation of $G_{x}$ on the tensor algebra 
$\otimes(T_xM)$ and its dual. On the space $S_{x} = T^*_xM\odot T^*_xM $ 
of symmetric $\left({}^{0}_{2}\right)$ tensors at $x$, the representation is 
$$
(\gamma,g_x) \to \mu^{*}_{\ss \gamma^{ -1}}g_x \ , \quad \gamma \in 
G_x\, , \ g_x \in  S_{x} \ .
\tag iso_rep
$$
Suppose $g \in \sg$, then by \(metric_symmetry) the linear isotropy 
representation \(iso_rep) on $g_x$ gives 
$$
\mu^{*}_{\ss \gamma^{\ss -1}}g_x = g_x  \ .
\tag iso_cond
$$
Thus if $g\in \sg$ then $g_x \in S_{x}^{\ss G_x}$, the fixed points of the 
linear isotropy representation.
Equation \(iso_cond) is the fundamental constraint on the values which 
any $g \in \sg$ can take and is the second type of reduction mentioned above.

The set $\Q$ is the space of sections of the bundle $Q$ whose fiber is 
$Q_x \subset T^*_xM \odot T^*_xM $, the set of Lorentz-signature 
quadratic forms on $T_{x}M$. Using \(iso_cond) we define the subset 
$K_x \subset Q_x$ to be
$$
K_{x} = Q_x^{\ss G_x} = \{ \ g_x \in  Q_x \quad | \quad 
\mu_{\ss\gamma^{\ss -1}}^* g_x = g_x \ , \quad \forall \gamma \in G_x \} \ .
$$ 
$K_x$ is the set of the Lorentz-signature quadratic forms on 
$T_xM$ which are invariant with respect to the action of $G_x$ on $Q_x$.  Now let 
$$
K = \bigcup_{\scriptscriptstyle x\in M}K_{x} \ . 
$$
We assume that $K\subset Q$ is a fixed rank sub-bundle of $Q$, 
then
by \(iso_cond) if $g \in \Qhat$ then $ g:M \to K$ is a section of 
$\pi\colon K \to M$.

Condition \(iso_cond) is only a 
necessary condition for $g \in \Qhat$, and so not every 
section of $K$ is a $G$-invariant tensor field. One more
step is required to determine the bundle over $M/G$ whose sections 
parameterize $\Qhat$. 
Since $G$ acts transversely to the fibers of $K$ \refto{Anderson2001}, 
the quotient $K/G$
is a bundle $\hat K\to M/G$ with the same  fiber as $K$ but with base 
space $M/G$. 
Each $G$-invariant metric $g\in \Qhat$ determines a section $\q$ of the bundle 
$\hat K$,
and each section $\q$ of $\hat K$ determines  a 
$G$-invariant metric $g\in \Qhat$ \refto{CGT2000}. Thus there exists a bijection,
$$
\Phi\colon \hat\Q \to \Qhat,
\tag phidef
$$
identifying the space $\Qhat$ of $G$-invariant metrics on $M$ with the space 
of sections, $\hat \Q$, of $\hat K$. 
Using the bijection $\Phi$, the set $\Qhat$ is thus parametrized by 
$m$ functions  of $r$ variables, $\q\in \hat\Q$, where $m$ is the dimension of the 
solution space to \(iso_cond) and $r=dim(M/G)$. 
We let $g(\q) = \Phi(\q)$ be the $G$-invariant metric defined by $\q$.\footnote*{For 
a large class of group actions the 
space $\hat{\cal Q}$ can be identified with the 
following set of Kaluza-Klein fields on $M/G$: (i) 
the space of metrics, (ii) 
a space of connections on a principal bundle, \ie non-Abelian gauge fields, 
and  (iii)
a space of scalar (``Higgs'') fields determining a non-linear sigma 
model. See \refto{Coquereaux1988} for 
details.}

Throughout we make the blanket hypothesis that a parallel construction arises for the parametrization of $G$-invariant tensor fields of any type. 
 In 
particular, we assume that 
there is a vector space of fields $\hat\s$ on $M/G$, 
which are sections of a vector bundle over $M/G$ and which parametrize 
the vector space $\sg$ of $G$-invariant symmetric $\left({}^{0}_{2}\right)$ 
tensor fields on $M$. The corresponding bijection is denoted by 
$$
P\colon \hat\s \to \sg,
\tag pdef
$$
where $P$ is a linear differential operator of order 0. We note that
$$
\Qhat\subset \sg, \quad \hat\Q\subset \hat\s,
$$
and
$$
\Phi = P\Big|_{\hat\Q}.
$$

Let $G$ act on $M$ and let $\Gamma$ be the Lie algebra of infinitesimal 
generators of the action of $G$. If $g \in \sg $ then $g$ satisfies the 
infinitesimal invariance equations
$$
L_{\ss X}g = 0, \qquad \forall \ X \in \Gamma\, , 
\tag Killing1
$$
which are the Killing equations if $g \in \Qhat$.  Suppose instead of being 
given $G$ that we are given a Lie algebra of vector-fields $\Gamma$ on 
an open set $U \subset M$.
In general $\Gamma$ will not be the infinitesimal generators of a Lie group 
acting on $M$ (or $U$)  so there is no corresponding notion of $G$-invariance. 
However, $G$-invariance can be replaced by $\Gamma$ invariance as follows.

Let  $\s^{\Gamma}$ be the symmetric $\left({}^{0}_{2}\right)$ tensor 
fields satisfying \(Killing1) on $U$ and similarly for $\Q^{\Gamma}\subset 
\s^{\Gamma}$. 
The parameterization of $\Q^\Gamma$ then proceeds along the same 
lines as $\Q^G$.  Let $\Gamma_{x_0} = \{ X \in \Gamma | X_{x_0} = 0\} $ 
be the isotropy algebra at $x_0 \in U$.
The condition 
$$
(L_{\ss X}g)_{x_0} = 0, \qquad X \in \Gamma _{x_0}
\tag infiso
$$
is the infinitesimal version of the isotropy constraint \(iso_cond). The solution
to \(infiso) determines the reduction of the metric components. The space
$M/G$ of reduced dependent variables is replaced by $U/\Gamma$ 
where points
on $U$ are identified if they lie on the same maximal integral manifold 
of $\Gamma$.
A functionally independent set of solutions to $ X(f(x)) = 0 \, , \ f \in C^\infty(U) $ 
form local coordinates on $U/\Gamma$.
If $\Gamma$ is given on a chart $(U,x^\alpha)$ and $ X = 
\xi^\alpha (x) \partial_{\alpha} \in \Gamma_{x_0}$ then 
$$
\left( \xi^\alpha_{,
\beta} \, g^{\beta\gamma} + \xi^\gamma_{,\beta}\,  
g^{\alpha\beta} \right)_{x_0} = 0 
$$
is the isotropy constraint \(infiso) which determines the admissible values 
of $g_{x_0}$ for $g \in \s^{\Gamma}$. 
Finally, we point out that if $G$ is connected then $\sg = \s^{\Gamma}$ 
where $\Gamma$ is the algebra of infinitesimal generators for the 
action of $G$ on $M$.

\taghead{3.}
\head{3. Lagrangians and Field Equations with Symmetry}

A Lagrangian for a metric field theory is a differential operator  
which, when applied to a metric $g$, yields a spacetime $n$-form 
$\lambda=\lambda(g)$.
Using the first variational formula we have
$$
\delta\lambda = E(\lambda)\cdot \delta g + d \eta(\delta g),
\tag deltaL
$$
where $\delta g\in \s$ is a symmetric tensor field, 
and $E(\lambda)$, which is the {\it Euler-Lagrange form of the 
Lagrangian $\lambda$}, is a symmetric $\left({}^{2}_{0}\right)$ 
tensor-valued $n$-form, and $\eta$ is a linear differential operator 
mapping any symmetric $\left({}^{0}_{2}\right)$ tensor field $h$ 
to an $(n-1)$-form $\eta(h)$. Both $E(\lambda)$ and $\eta(\delta g)$ are 
local functions of the metric and its derivatives. Note that the 
$(n-1)$-form $\eta(\delta g)$ 
is only defined up 
to the addition of an exact $(n-2)$ form locally constructed from the 
metric, the metric variation, and their derivatives. (The metric 
variation can only appear linearly.)
The {\it Euler-Lagrange equations} (or the {\it field 
equations}) are the system of differential equations
$$
E(\lambda) = 0.
\tag ELeqns
$$

In coordinates $x^{\alpha}$ on $M$ we have
$$
\lambda = L\nu,
\tag Lag_density
$$
where
$$
\nu = dx^{1}\wedge dx^{2}\wedge\cdots \wedge dx^{n},
\tag fixedVolumeForm
$$
and 
$$
L=L(x^{\alpha},g_{\mu\nu},g_{\mu\nu,\sigma},
\ldots,g_{\mu\nu,\sigma_{1}\cdots\sigma_{k}})
$$ 
is a local function of the metric and its derivatives  called  the 
{\it Lagrangian density}. We have
$$
E(\lambda) = E^{\alpha\beta}(g) 
{\partial\over\partial x^{\alpha}}\otimes {\partial\over\partial 
x^{\beta}}
\otimes \nu,
$$
with
$$
E^{\alpha\beta}=E^{\beta\alpha} = {\partial L\over\partial 
g_{\alpha\beta}} - \partial_{\sigma}{\partial L\over\partial 
g_{\alpha\beta,\sigma}} + \partial_{\sigma_{1}}\partial_{\sigma_{2}}
{\partial L\over\partial 
g_{\alpha\beta,\sigma_{1}\sigma_{2}}} +\cdots + (-1)^{k}
\partial_{\sigma_{1}}\cdots \partial_{\sigma_{k}}
{\partial L\over\partial 
g_{\alpha\beta,\sigma_{1}\cdots \sigma_{k}}},
$$
$$
E(\lambda)\cdot \delta g\ = E^{\alpha\beta}\delta g_{\alpha\beta}\, \nu,
$$
and
$$
\eta(\delta g) = \left(\eta^{\alpha\beta}\delta g_{\alpha\beta} + 
\eta^{\alpha\beta,\gamma}
\delta g_{\alpha\beta,\gamma} 
+\cdots + \eta^{\alpha\beta,\gamma_{1}\cdots \gamma_{k-1}}\delta 
g_{\alpha\beta,\gamma_{1}\cdots \gamma_{k-1}}
\right)
dx^{1}\wedge 
dx^{2}\wedge\cdots \wedge dx^{n-1}.
$$

For example, the Lagrangian we shall use repeatedly in \S6 is 
the Einstein-Hilbert Lagrangian, which takes the form
$$
\lambda(g) = R(g)\epsilon(g),
\tag EH
$$
where $R$ is the scalar curvature of the metric $g$ and $\epsilon$ is 
the volume form of $g$. The Euler-Lagrange form is then 
given by
$$
E(\lambda) = -\G\otimes \epsilon = -\sqrt{|g|}
\G^{\alpha\beta}\partial_{\alpha}\otimes\partial_{\beta}\otimes \nu,
\tag EHeq
$$
where $\G$ is the $\left({}^{2}_{0}\right)$ form of the Einstein 
tensor. The boundary term $\eta$ can be chosen to be \refto{Iyer1994}
$$
\eta(h) =-*[{\rm div}h - d({\rm tr}h)]= -\epsilon_{\alpha 
\beta\gamma}{}^{\delta}\left(\nabla^{\gamma}h_{\delta\gamma} - 
\nabla_{\delta}h_{\gamma}^{\gamma}\right)dx^{\alpha}\wedge 
dx^{\beta}\wedge dx^{\gamma},
$$
where $\nabla$ is the Levi-Civita connection defined by $g$.

We say that a gravitational theory is {\it 
$G$-invariant} if its 
Lagrangian $\lambda$ 
is {\it $G$-equivariant} as a map from $\Q$ to the set of $n$-forms 
on $M$: 
$$
\lambda(\mu^{*}_{\gamma} g) = \mu^{*}_{\gamma}\lambda(g),\quad
\forall\ \gamma\in G,
\tag LEquivariance
$$
that is, $\lambda$ is suitably ``covariant'' with respect to the symmetry group 
action. Granted the $G$-equivariance of the Lagrangian, the naturality of the 
Euler-Lagrange form \refto{Olver1993} implies that it
is also 
$G$-equivariant:
$$
E(\lambda)(\mu^{*}_{\gamma}g) = \mu^{*}_{\gamma}E(\lambda)(g),
\quad
\forall\ \gamma\in G.
$$

The $G$-equivariance of the Lagrangian and Euler-Lagrange form  
guarantees the following key fact:  the Lagrangian and the Euler-Lagrange form 
are $G$-invariant tensor fields when 
evaluated on a $G$-invariant metric $g\in \Qhat$:
$$
\left.\eqalign{
\mu_{\gamma}^{*}\lambda(g) &= \lambda(g),\cr 
\mu_{\gamma}^{*}E(\lambda)(g) &= E(\lambda)(g).}
\right\}\forall\ \gamma\in G,\ g\in \Qhat
\tag invariants
$$

We shall in all that follows assume that the field theory under 
consideration is $G$-invariant.
In general relativity and its variants 
it is normally assumed that the Lagrangian and field equations are in 
fact equivariant 
with respect to  the whole spacetime diffeomorphism group, 
that is, the theory is ``generally covariant''. In this case, for any 
symmetry reduction the 
$G$-equivariance of the Lagrangian and field equations is  
guaranteed since the symmetry group is always acting as a subgroup of the 
spacetime diffeomorphism group.  Nonetheless, we shall not need to 
assume general 
covariance in what follows. Thus our results can be applied to  all 
generally covariant (metric) theories of gravity, but also to theories 
which are not covariant because, \eg they  involve fixed, background fields. 
 
Granted a $G$-equivariant Lagrangian, if we extend the $G$-action to the 
field variations 
$\delta g\in \s$, then it follows that $\delta \lambda$ is $G$-equivariant:
$$
\delta\lambda(\mu^{*}_{\gamma}g,\mu^{*}_{\gamma}\delta g)
=\mu^{*}_{\gamma}\delta\lambda(g,\delta g),\quad \forall\ \gamma\in G.
$$
Because of the $G$-equivariance of the Euler-Lagrange expression it 
then follows that the volume term $E(\lambda)\cdot \delta g$ and 
the boundary term $d\eta(\delta g)$ are separately 
$G$-equivariant.  However, it is not immediately clear that the 
$(n-1)$-form $\eta(\delta g)$ featuring in the boundary term can be 
chosen to be 
$G$-equivariant. Iyer and Wald \refto{Iyer1994} have established the existence of a 
$G$-equivariant choice of $\eta$  in any generally covariant 
field theory. More generally, it is possible to establish the existence of a 
$G$-equivariant choice of $\eta$ for any $G$-invariant field theory 
admitting a 
$G$-invariant metric \refto{AFT}. We assume that such a choice of $\eta$ has 
been made so that 
$$
\eta(\mu^{*}_{\gamma}g,\mu^{*}_{\gamma}\delta g) = 
\mu^{*}_{\gamma}\eta(g,\delta g),\quad \forall\ \gamma\in G.
$$
Then we have that for any $G$-invariant metric $g\in \Qhat$ and 
$G$-invariant metric variation $\delta g\in \sg$:
$$
\left.\eqalign{
\mu_{\gamma}^{*}\delta \lambda&= \delta\lambda,\cr 
\mu^{*}_{\gamma}\eta &= \eta}\right\}\quad \forall\ 
\gamma\in G,\ g\in \Qhat,\ \delta g \in \sg.
$$

We conclude this section by reminding the reader that  the 
foregoing discussion admits a purely local, infinitesimal 
description. One simply replaces the group $G$ acting on $M$ 
with an algebra of vector fields $\Gamma$ on $U\subset M$ and uses the 
corresponding notions of infinitesimal invariance/equivariance. 
\bigskip

\taghead{4.}
\head{4. Symmetry Reduction of the Field Equations and Lagrangian}

Given a system of field equations and a  group action $(G,\mu)$, we can ask for 
solutions to the field equations that are $G$-invariant. This simply 
means that we ask for metrics $g$ that satisfy
$$
E(\lambda(g)) = 0,\quad \mu_{\gamma}^* g = g,\quad \forall\ \gamma\in G,
$$
that is, we are restricting the field equations from $\Q$ to $\Qhat$:
$$
E(\lambda)(g(\q)) = 0.
$$
 The principal idea behind the theory of 
symmetry reduction is that a system of $G$-equivariant field 
equations, when restricted to $G$-invariant metrics,
is equivalent\footnote*{See Appendix B for an explanation of how we 
are defining differential equations to be ``equivalent''.} 
to a 
simpler system of {\it reduced field equations}
$
\hat\Delta(\q) = 0
$ 
for the fields $\q$ on $M/G$ that parametrize 
$\Qhat$ (\cf \S2) \refto{CGT2000}:
$$
E(\lambda)(g(\q)) = 0\ {\rm on}\ M
\quad \Longleftrightarrow \quad \hat\Delta(\q) = 0\ {\rm on}\ M/G.
\tag reducedeq
$$
Typically, the equations $\hat\Delta=0$ 
are considerably more tractable than the original field 
equations since the number of independent and/or dependent variables 
has been reduced.

One 
important feature of the reduction from 
$E(\lambda)=0$ to $\hat \Delta=0$ is that,  just as 
the isotropy group (if any) leads to a reduction in the number of 
independent components of the $G$-invariant metrics, it also reduces 
the number of independent field equations. This occurs because  
the field equation tensor 
$E(\lambda)$, upon 
restriction to an element of $\Qhat$, becomes a $G$-invariant symmetric 
$\left({}^{2}_{0}\right)$ tensor-valued $n$-form, $E(\lambda)(g(\q))$, 
on $M$ and it obeys isotropy 
constraints at each point in the same way as was discussed in \S 2 for 
the metric. Therefore, provided the group 
action is not free, the number of reduced field equations $\hat\Delta=0$ 
will be less 
than the original number of field equations, just as the number of 
variables $\hat q$ will be less than the number of components of a 
generic spacetime metric.

As with the field equations, the Lagrangian $\lambda(g)$ 
when restricted from $\Q$ to $\Qhat$ determines a 
(much simpler)
{\it reduced Lagrangian} $\hat\lambda(\q)$ for the fields $\q$.
The construction of the reduced Lagrangian is slightly more 
complicated than the reduced field equations because 
the original Lagrangian is an $n$-form on $M$ and the reduced 
Lagrangian must be an $(n-l)$-form on 
$M/G$, where $l$ is the dimension of the group orbits in $M$. 
Let us therefore spell out the definition of the reduced Lagrangian 
in some detail.

We
define a {\it $G$-invariant $l$-chain} $\chi$ on $M$ to be a totally antisymmetric 
tensor of type $\left({}^{l}_{0}\right)$ which  (1) is a $G$-invariant 
tensor field: 
$$
\mu^{*}_{\gamma}\chi = \chi,\quad \forall\ \gamma\in G,
$$
and (2) is  everywhere 
tangent to the orbits of $G$ on $M$. This last requirement means that 
at any point $x\in M$ we have
$$
\chi = \chi^{i_1\cdots i_{l}}X_{i_{1}}\cdots 
X_{i_{l}},
$$
where 
$(
X_{i_{1}},X_{i_{2}},
\ldots,X_{i_{l}})
$
are a linearly independent set of generators of the isometry group action 
(Killing vector fields) at $x$.
Suppose that 
$\omega=\omega(g)$ is a $G$-equivariant $p$-form locally constructed from the 
metric and its derivatives: 
$$
 \omega(\mu^{*}_{\gamma}g) = \mu^{*}_{\gamma}\omega(g),\quad 
\forall\ \gamma\in G.
 $$
Consider the $(p-l)$-form $\chi\hook\omega(g(\q))$ obtained from $\omega$ 
by contraction 
with $\chi$ and evaluation on a $G$-invariant metric $g(\q)\in \Qhat$. 
This form is 
$G$-invariant and
satisfies
$$
X\hook \{\chi\hook \omega(g(\q))\} = 0, 
$$
with $X$ being any infinitesimal generator of the group action $\mu$ on 
$M$.\footnote*{Such forms are  called {\it basic}.} This 
means that there exists a unique $(p-l)$-form $\hat 
\omega(\q)$ on $M/G$ locally constructed from the fields $\q$ and 
their derivatives  such that
$$
\chi\hook\omega(g(\q)) = \pi^{*}\hat \omega(\q),
\tag rhodef
$$
where $\pi\colon M\to M/G$ is the  projection from $M$ to its 
quotient by the $G$ action.
So, given a choice of $G$-invariant $l$-chain on $M$, we have a map, 
$$
\rho_{\chi}\colon (\Omega^{p}(M))^{\ss G}\to \Omega^{p-l}(M/G),
$$
that associates a $(p-l)$ form $\hat\omega(\q)$ on $M/G$, 
to every $G$-invariant spacetime $p$-form $\omega(g(\q))$:
$$
\rho_{\chi}(\omega(g(\q)))=\hat\omega(\q) . 
$$
We call $\rho_{\chi}$ the {\it reduction map} associated to the 
$l$-chain $\chi$.

If we restrict the Lagrangian to $\Qhat$, 
the resulting form $\lambda(g(\q))$ is $G$-invariant.
We can now apply the reduction map
to obtain the {\it reduced Lagrangian} $\hat \lambda(\q)$:
$$
\hat \lambda(\q) = \rho_{\chi}(\lambda(g(\q))).
\tag lhat
$$
Associated to this reduced Lagrangian is a system of  
Euler-Lagrange equations $E(\hat\lambda) = 0$, for the fields $\q$
on $M/G$,  
where $E(\hat\lambda)$ is defined via the first variational formula
$$
\delta \hat\lambda = E(\hat\lambda)\cdot \delta \q + d\hat \eta( 
\delta \q),
$$
with $\delta\q\in \hat\s$.
We emphasize that the reduced Lagrangian and its Euler-Lagrange form 
depend upon the choice made for the invariant chain $\chi$.

The construction  of the reduction map and the
reduced Lagrangian goes through 
when $\chi$ is only defined on a $G$-invariant 
open set $U\subset M$. In this case, the reduction map defines forms,
\eg the
reduced Lagrangian $\hat\lambda$, 
only on $V=\pi(U)\subset M/G$.  All of our preceding and subsequent 
considerations will apply in this setting provided we replace $M$ by $U$ and 
$M/G$ by $V=U/G$. And, as usual, we remind the reader that  the 
foregoing discussion admits a purely local, infinitesimal 
description. One simply replaces the group $G$ acting on $M$ 
with an algebra of vector fields $\Gamma$ on an open set $U\subset M$ 
and uses the 
corresponding notions of infinitesimal invariance/equivariance. In this setting, 
the reduced Lagrangian and field equations are obtained by restricting to 
$\s^{\ss \Gamma}$ and proceeding as before.

\taghead{5.}
\head{5. The Principle of Symmetric Criticality}

The principle of symmetric criticality  can now be stated in 
terms of the ingredients described in the previous sections.

\proclaim Definition 5.1.
A group action  obeys the {\bf Principle of Symmetric 
Criticality (PSC)} if about each $x\in M$ there exists a $G$-invariant open 
neighborhood $U$ and a $G$-invariant chain on $U$   
 such that, for any $G$-equivariant Lagrangian 
$\lambda$, 
the reduced field equations are equivalent (in the sense of Appendix B)
to the 
Euler-Lagrange equations of the reduced Lagrangian $\hat\lambda$,
$$
E(\lambda)(g(\q)) = 0 \Longleftrightarrow
E(\hat\lambda)(\q) = 0.
\tag ELpsc
$$

We emphasize that PSC is 
a property of a group action and not a property of a specific 
Lagrangian. It is possible to have a particular Lagrangian that yields a correct 
reduced Lagrangian for some symmetry reduction even if PSC fails in 
the sense that not all Lagrangians can be successfully reduced. As an 
extreme 
example, the Lagrangian $\lambda=0$ 
will always reduce to give the correct field equations even 
if PSC is not valid for the chosen group action.  The point of 
PSC, when it is valid, is that it guarantees the  reduced 
Lagrangian 
is correct irrespective of the starting Lagrangian. 

PSC 
guarantees that 
solutions to 
$E(\hat\lambda)=0$ are in one-to-one correspondence with 
$G$-invariant solutions to $E(\lambda)=0$.\footnote\dag{Of course, for a 
randomly chosen Lagrangian 
there is no guarantee that the various differential equations admit 
any solutions at all!  The notion of 
equivalence  we have adopted for defining PSC (see Appendix B) is particularly 
convenient in this regard since it only relies upon the relation 
between differential equations, not on any properties of their solution 
spaces --- not to mention that such properties may be difficult to 
establish. }
This happens, for example, with 
the spherically symmetric reductions mentioned in \S 1.
However, as pointed out in \S 1, PSC need not be valid. 
 As we shall see, there are exactly two 
obstructions to the validity of PSC in relativistic field theory 
and they can be characterized 
quite explicitly in terms of properties of the symmetry group action 
$(G,\mu)$.

In order to obtain conditions for the validity of PSC, 
we want to compare the Euler-Lagrange form $E(\lambda)(g(\q))$ 
with the Euler-Lagrange form of the reduced Lagrangian, 
$E(\hat\lambda)(\q)$. To this end, we note that the 
reduction map $\rho_{\chi}$, since it is defined independently of the metric, 
commutes with the process of field 
variation so that
$$
\eqalign{
\delta \hat\lambda  &= \rho_{\chi}(\delta_{\q}\lambda(g(\q)))\cr
&=\rho_{\chi}\bigg(E(\lambda)(g(\q))\cdot\delta_{\q} g(\q) + d\eta( 
\delta_{\q} g(\q))\bigg).}
\tag delta
$$
In \(delta) we have  introduced the notation $\delta_{\q}$, which indicates a 
variation in the fields $\q$ on $M/G$. 
In particular,
$$
\delta_{\q}g = P\cdot\delta\q,
\tag deltag
$$
where $P$ is the zeroth order linear differential operator defined in \(pdef).
Thus, $P$  is playing the role of the differential of the 
map $\Phi$ in \(phidef) and  $\delta_{\q}g$ represents a  
tangent vector at $g(\q)$ to $\Qhat$. 

As noted in \S4, $\lambda$, $E(\lambda)$ and $\eta$ are $G$-equivariant.  
This means that 
$E(\lambda)(g(\q))\cdot\delta_{\q} g(\q)$ and $d\eta( 
\delta_{\q} g(\q))$ are  $G$-invariant forms on $M$. 
We can therefore 
apply $\rho_{\chi}$ to each term on the right-hand side of \(delta). 
Comparing   \(delta) with \(lhat) we have
$$
E(\hat\lambda)\cdot  \delta \q + d\hat \eta( 
\delta \q)
=\rho_{\chi}\bigg(E(\lambda)(g(\q))\cdot (P\cdot\delta\q)\bigg) +
\rho_{\chi}\bigg(d\eta( 
P\cdot\delta\q)\bigg).
\tag deltadeltahat
$$
Comparing this with \(ELpsc),
we have the following conditions for the validity of PSC.

\proclaim Theorem 5.2.
The following two conditions are necessary and sufficient for the 
validity of PSC.
\medskip\noindent
PSC1: About each $x\in M$ there  exists a $G$-invariant open 
neighborhood $U$ and a $G$-invariant chain $\chi$ on $U$   such that
for each $G$-invariant $(n-1)$-form $\eta$ on $U$ there 
exists an $(n-l-1)$-form $\hat\eta$ on $U/G$ with
$$
\rho_{\chi}(d\eta) = d\hat\eta.
\tag coch
$$
Here $l$ is the dimension of the group orbits in $M$.
\bigskip\noindent
PSC2:  For all $G$-equivariant Lagrangians $\lambda$
$$
E(\lambda)(g(\q))\cdot P =0 
\quad \Longleftrightarrow\quad
E(\lambda)(g(\q))=0.
\tag psc2
$$

\proof
These conditions can be seen to be sufficient for PSC as follows. 
Eq. \(coch) in PSC1 implies that 
$$
\rho_{\chi}\bigg(d\eta( 
P\cdot\delta\q)\bigg)=d\hat \eta( 
\delta \q)
$$
in \(deltadeltahat). 
Therefore we  have  that $E(\hat\lambda)$ 
is determined by the  
``volume term'' on the 
right-hand side of \(deltadeltahat):
$$
E(\hat\lambda)\cdot \delta\q = 
\rho_{\chi}\bigg(E(\lambda)(g(\q))\cdot(P\cdot\delta\q)\bigg)\quad 
\forall\ \delta\q.
\tag volterm
$$
Since $\rho_{\chi}\colon(\Omega^{n}(U))^{\ss G}\to \Omega^{n-l}(U/G)$ 
is an isomorphism, it follows from \(volterm) 
that the Euler-Lagrange equations of the reduced Lagrangian 
are always
equivalent to at
least a subset of the correct reduced field equations, namely, 
$$
E(\hat\lambda)=0\Longleftrightarrow 
E(\lambda)(g(\q))\cdot P=0.
\tag EdotP
$$
Equation \(psc2) of PSC2 then  reduces  \(EdotP)  to \(ELpsc). 
Conditions PSC1 and PSC2 are   necessary 
for PSC since PSC asserts that \(ELpsc) holds for {\it all} $G$-equivariant 
Lagrangians. If PSC1 or PSC2 is not satisfied there 
will be {\it some} Lagrangian for which the reduced Lagrangian will 
not yield valid Euler-Lagrange equations. We will exhibit 
such PSC-violating Lagrangians to \S 5.1 and \S 5.2. \square
  
 The role of PSC1 is to guarantee that 
the restriction of the 
boundary term to $\Qhat$ does not introduce any additional ``volume terms'' 
proportional to $\delta\q$ on the right hand side of \(deltadeltahat).
Such terms will have the effect of supplying 
erroneous contributions  
to the Euler-Lagrange equations of the reduced Lagrangian, rendering 
these equations incorrect.
 Thus PSC1 guarantees 
that the field equations produced by the reduced Lagrangian correctly 
yield at least 
a subset of the reduced field equations.  We note that the Bianchi class B models 
mentioned in \S  1 do not satisfy PSC1, and this is the reason PSC fails in these models.
 Granted 
 PSC1,
 the role of PSC2 is to guarantee that the reduced 
Lagrangian supplies {\it all} of the reduced field equations.  

In the next two subsections we provide pointwise conditions on the group 
action which are necessary and sufficient for  PSC1 and PSC2. In the third 
subsection a number of special situations are considered where the validity 
of  these pointwise conditions can be established.

\subhead{5.1 PSC1: The Lie algebra condition}

 PSC1  
has already been studied in \refto{Anderson1997} under 
the hypotheses of (i) the existence of a $G$-equivariant boundary term $\eta$ 
in the variational 
principle \(deltaL), (ii) a transverse symmetry group action,  
 and (iii) connected $G$ and $G_{x}$. 
As mentioned in \S 4, 
in general relativity we always have a $G$-equivariant boundary term. 
Moreover, it is straightforward to check that 
the analysis of PSC1 appearing in 
\refto{Anderson1997} does 
not, in fact, depend upon the transversality assumption and that   
these results immediately generalize 
to allow for a disconnected symmetry group and/or  isotropy subgroups. 
So, referring to \refto{Anderson1997} for proofs where necessary, 
we have the following results.

\proclaim Proposition 5.3.  PSC1 is equivalent to the condition that 
about each point $x\in M$ there is a $G$-invariant open neighborhood 
$U$ and 
 $G$-invariant chain $\chi$ on $U$ such that
$$
\rho_{\chi}(d\eta) = d\rho_{\chi}(\eta),
\tag cochain
$$
for all $G$-invariant differential forms $\eta$ on $U$.

\proof: See \refto{Anderson1997}. \square

Therefore, following \refto{Anderson1997}, if a $G$-invariant chain $\chi$ 
 exists on a $G$-invariant open set $U$,
 so that PSC1 holds in $U$, we say that $\chi$ defines a {\it 
 cochain map} $\rho_\chi\colon (\tilde\Omega^{*}(U))^{G}\to 
 \Omega^{*-l}(U/G)$, from the space of $G$-basic forms on $U$ to 
 the corresponding forms on $U/G$. PSC1 requires that a cochain map 
 exist in a $G$-invariant neighborhood of any point $x\in M$.
 Necessary and sufficient conditions for the existence of a cochain 
map in a $G$-invariant open set, 
expressed in terms of properties of the group 
action,  are as follows. 

\proclaim Proposition 5.4. There exists a 
cochain map $\rho_\chi$ satisfying \(cochain) on a $G$-invariant 
open set $U$ 
if and only if there  exists on  $U$ a $G$-invariant 
chain that has vanishing Lie derivative along all
$G$-invariant vector fields  on $U$. 

\proof: See \refto{Anderson1997}. \square

\bigskip

The  condition stated in   
Proposition 5.4 for the existence of a cochain map can be 
reformulated in terms of the relative Lie algebra 
cohomology ${\cal H}^{*}(\Gamma,G_{x})$ of the Lie algebra $\Gamma$ 
of $G$
relative to its isotropy subgroups $G_{x}$. 
The cohomology ${\cal H}^{l}(\Gamma,G_{x})$ at degree $l$ is 
defined as follows. Fix a basis $e_{a}$, $a=1,2,\ldots, {\rm dim}G$ 
for the Lie algebra. The structure constants of the Lie algebra are 
then
defined by the Lie bracket:
$$
[e_{a},e_{b}]=C_{ab}{}^{c} e_{c}.
$$
The Lie algebra $\Gamma$ can  be viewed as the space of left-invariant 
vector fields on the manifold $G$ and the Lie bracket as the vector 
field commutator. The dual basis of left-invariant 1-forms on $G$, 
$\omega^a$, $ a=1,2,\ldots, {\rm dim}G$,
satisfy
$$
d\omega^{a} =- \half C_{bc}{}^{a} \omega^{b}\wedge\omega^{c}.
\tag Lie_d
$$
The Lie algebra cohomology of $G$ at degree $l$ is defined as the 
space of closed $l$-forms on $G$ modulo the exact 
$l$-forms, with all forms being left-$G$-invariant. It can be computed
using the exterior derivative formula \(Lie_d).  
The {\it relative} Lie algebra cohomology
${\cal H}^{l}(\Gamma,G_{x})$ is defined as
the set of closed 
$l$-forms on the group $G$ modulo 
the set of exact $l$ forms, where all forms are left-$G$-invariant and 
right $G_{x}$-basic. This last condition means 
that, with 
$\Gamma_{x}=Lie(G_{x})$, all forms
are required to be invariant under the right action of $G_{x}$ on $G$ and 
to satisfy
$$
X\hook\omega=0,\quad \forall X\in \Gamma_{x}.
$$
A $G_{x}$-basic form on $G$ is the pull-back of a form on $G/G_{x}$.
The relative Lie algebra 
cohomology ${\cal H}^{*}(\Gamma,G_{x})$ computes the  $G$-invariant 
de Rham 
cohomology  of the orbit $G/G_{x}$ through $x\in M$. 

Let $l$ be the dimension of the group orbits in $M$. Then we have the 
following necessary and sufficient condition for PSC1.

\proclaim Proposition 5.5. 
Around each point $x\in M$ there is a $G$-invariant 
neighborhood $U$ such that a
cochain map exists on $U$ if and only if 
$$
{\cal H}^{l}(\Gamma,G_{x}) \neq 0,\quad \forall x\in M.
\tag Lie
$$

\proof See \refto{Anderson1997} for a proof of existence of a cochain 
map in a neighborhood of each 
point $x\in M$ when \(Lie) holds. 
It is straightforward to check that this neighborhood 
can always be chosen to be $G$-invariant. \square

We call \(Lie) the {\it Lie algebra condition} for PSC. 
 If $x$ and 
$x^{\prime}$ lie in the same $G$ orbit then ${\cal H}^{*}(\Gamma,G_{x})$ is 
isomorphic to ${\cal H}^{*}(\Gamma,G_{x^{\prime}})$,  so one need only check the 
Lie algebra condition along a cross section of the group action.

Finally, we note that Proposition 5.4 can be used to show the necessity of
PSC1 in Theorem 5.2. Suppose 
that PSC1 does not hold, then there is no cochain map and according to 
Proposition 5.4
there will 
exist a $G$-invariant vector field $S$ such that $L_{\ss S}\chi\neq0$ 
for any $G$-invariant chain $\chi$. In fact, as shown in 
\refto{Anderson1997}, $S$ can be chosen tangent to the group orbits 
and there exists a smooth, 
non-trivial $f$  
such that
$$
L_{\ss S}\chi = f\chi,
\tag lschi
$$
where both $S$ and $f$ can be chosen 
independently of the $G$-invariant $l$-chain 
$\chi$.
Consider the trivial Lagrangian 
$$
\eqalign{
\lambda(g) &= d (S \hook \epsilon(g))
\cr
&= L_{\ss S} \epsilon(g) \cr 
&  = \half g^{\mu\nu}( L_{\ss S} g_{\mu\nu}) \epsilon(g),}
\tag trivialL
$$ 
where $\epsilon(g)$ is the volume form of $g$.
Being exact, this Lagrangian has identically vanishing Euler-Lagrange form. 
On the other hand, if we compute the pull back of the reduced 
Lagrangian $\hat\lambda$ to $M$, we have
$$
\eqalign{
\pi^{*}\hat\lambda(\q)&=\chi \hook \lambda(g(\q))\cr 
&= L_{\ss S} (\chi \hook \epsilon (g(\q))) -
(L_{\ss S} \chi) \hook \epsilon (g(\q)) \ .}
\tag reducedtrivial
$$  
The first term in \(reducedtrivial) is zero because 
$\chi\hook\epsilon(g(\q))$ is basic and of  degree
$(n-l)$. Therefore
$$
\pi^{*}\hat\lambda(\q)= - f \chi \hook \epsilon(g(\q)),
$$
which implies that the reduced Lagrangian is of order zero in $\q$ 
and is non-trivial provided the space of invariant metrics is 
non-trivial. 
Therefore $\hat\lambda$ has non-vanishing
Euler-Lagrange form, which violates PSC.

\subhead{5.2 PSC2: The Palais condition}

For the derivation of the results of this section and their subsequent 
application to various examples in \S6, it is  convenient 
to define the symmetric $\left({}^{2}_{0}\right)$ tensor 
field
$\Delta(g)\in \s^*$ by
$$
E(\lambda(g))  = \Delta(g)\otimes \epsilon(g),
\tag efe
$$
where $\epsilon(g)$ is the volume form of the metric $g$. 
For example, in \(EHeq) where 
$\lambda(g)$ is the Einstein-Hilbert Lagrangian form,  $\Delta(g)$ is 
minus the Einstein tensor.  
Note that 
$\Delta(g)$
is $G$-equivariant if $\lambda(g)$ is. This implies that $\Delta(g(\q))$ is a 
$G$-invariant symmetric 
$\left({}^{2}_{0}\right)$ tensor field, $\Delta(g(\q))\in 
(\s^{*})^{\ss G}$. 
From \(efe),  PSC2 is  equivalent to the 
statement that,  for all 
field equations coming from $G$-equivariant Lagrangians,
$$
\Delta(g(\q))\cdot P = 0 \Longleftrightarrow \Delta(g(\q)) = 0.
\tag scp2eqalt
$$

To describe necessary and sufficient conditions for \(scp2eqalt) to be 
satisfied, we recall some definitions from \S2. Let $S_x=T_{x}^{*}M\odot 
T_{x}^{*}M$ be  the set of
possible values of symmetric $\left({}^{0}_{2}\right)$ 
tensor fields at the point $x\in M$ and denote by 
$\sgx\subset S_x$ the 
vector space of $G_{x}$-invariant symmetric $\left({}^{0}_{2}\right)$ 
tensors at $x\in M$.  At each $x\in M$,
$
\delta_{\q} 
g(\q)_{x}\in \sgx.
$
Similarly, we 
have the dual representation  of the group $G_{x}$ 
acting on $S^{*}_{x}=T_{x}M\odot T_{x}M$ and we denote the fixed 
points of this action as 
$\sgstarx$, so that $\sgstarx\subset S^{*}_{x}$ is the vector space of 
$G_{x}$-invariant $\left({}^{2}_{0}\right)$ tensors at $x\in M$. 
By equivariance of $\Delta$, at each point $x\in M$ we have that 
$$
\Delta(g(\q))_{x} \in \sgstarx.
\tag ELvertG
$$ 
Finally, let us denote by $\sgxo\subset  S^{*}_{x}$ the annihilator of 
$\sgx$. $\sgxo$  is the set of elements 
$\omega\in S^{*}_{x}$ such that $\omega\cdot h= 0$ for all $h\in 
\sgx$.

\proclaim Proposition 5.6.
A necessary and sufficient condition for the 
validity of \(scp2eqalt) -- and hence PSC2 -- is that 
$$
\sgstarx \cap \sgxo = 0, \quad\forall\ x\in M.
\tag intersection
$$

\proof
Recall that $\hat \s$ is the space of sections of a vector bundle 
over $M/G$ that parametrizes $\sg$. Let us denote by $\hat\s^{*}$ 
the space of sections of the dual vector bundle. 
Define the linear mapping 
$$
H\colon (\s^{*})^{\ss G}
\to \hat\s^{*}
$$
by
$$
H(v) = v\cdot P,\quad  v\in (\s^{*})^{\ss G},
$$
so that
$$
\Delta(g(\q))\cdot P = H \bigg(\Delta(g(\q))\bigg).
$$
Because 
$$
P\colon \hat\s\to \sg
$$ 
is an isomorphism,
it follows that $H$ is surjective.
Therefore
to prove sufficiency of \(intersection) for PSC2 (in the form 
\(scp2eqalt)) it is enough to show 
that $H$ is injective when \(intersection) holds. We have that
$$
{\rm ker}(H) =\{v\in (\s^{*})^{\ss G}|v\cdot w =0,\ \forall\ w\in 
\s^{\ss G}\}.
$$
Now, if $v\in(\s^{*})^{\ss G}$ then at each $x\in M$ we have that 
$v_{x}\in \sgstarx$. 
Our regularity assumptions on the group action guarantee that for 
each element $w_{x}\in \sgx$ there is a smooth section $w\in \sg$ 
taking the value $w_{x}$ at $x\in M$. Therefore, if $v\in {\rm ker}(H)$ 
then $v_{x}\in \sgstarx \cap \sgxo$
so that condition \(intersection) forces ${\rm ker}(H)=0$, \ie  
\(intersection) implies that $H$ is 
an isomorphism and PSC2 holds.

To verify necessity of \(intersection) for PSC2, fix 
$h_{x}\in \sgstarx\cap \sgxo$ and 
let $h\in \s^{*}$  be any 
$G$-invariant symmetric 
$\left({}^{2}_{0}\right)$ tensor field on $M$ that takes the value $h_{x}$ 
at $x$. Such a form exists by virtue of the regularity assumptions we 
have made on the group action \refto{CGT2000, Anderson2001}.  Let 
$\omega$ be a $G$-invariant volume form on $M$. Such a form always 
exists since we assume the existence of a $G$-invariant metric. Consider the 
$G$-equivariant Lagrangian
$$
\lambda =h^{\alpha\beta}g_{\alpha\beta}\, \omega.
\tag badL
$$
PSC2 requires that
$$
h\cdot P = 0,\quad\Longleftrightarrow\quad
h=0.\tag intersectlag
$$
At the point $x$, \(intersectlag) implies that $h_{x}=0$. 
Therefore, \(intersection) is necessary and sufficient for 
PSC2 to hold. \square

In the context of PSC for $G$-invariant functionals on Banach manifolds Palais has arrived at condition \(intersection) \refto{Palais1979}, but with $G$ acting on $\s$ rather than $G_x$ acting on $S_x$ as we have here.  We therefore refer to \(intersection) as the 
{\it Palais condition} 
for PSC. Note, however, that unlike the case 
 in
 \refto{Palais1979},  condition \(intersection) is necessary but not 
 sufficient for PSC since \(intersection) only guarantees PSC2. 
It is clear that the Palais condition can only fail when (1) the group 
action is not free, 
\ie there is non-trivial isotropy, and (2) the isotropy group is 
represented non-trivially on the space of values of the fields at 
each point. Conditions (1) and (2) imply the action of the symmetry 
group on the bundle of fields is necessarily non-transverse. So, for 
transverse group actions (such as considered in \refto{Anderson1997}), 
PSC2 is always satisfied.

Finally, we note that equation \(badL) is an example of a $G$-equivariant
Lagrangian that violates PSC when PSC2 does not hold, thus 
establishing the necessity of PSC2 in Theorem 5.2. 

\subhead{5.3 Further developments, simplifications and specializations}

Here we provide some additional results on the Lie algebra condition \(Lie) 
and the Palais condition \(intersection) for PSC, 
given that the group actions of interest preserve a spacetime metric and/or 
granted additional simplifying assumptions.
We begin by characterizing the Palais condition in terms of 
inner products on $\sgx$
and then show that the Palais condition is always satisfied  if there is a 
$G$-invariant Riemannian metric on $M$.  We then assume that $G$ and $G_{x}$ are 
connected so that the Lie algebra and Palais conditions for PSC can be 
characterized using 
infinitesimal methods.

\proclaim Proposition 5.7.
Condition \(intersection) holds  if and 
only if every $G_{x}$ 
invariant metric on $S_x$ is non-degenerate when restricted to 
$\sgx$.

\proof
Let $B\colon S_x\times S_x\to \reals$ be a 
$G_{x}$-invariant, non-degenerate quadratic form on the space of 
symmetric $\left({}^{0}_{2}\right)$ tensors at $x\in M$. This means that if 
$h$ and $k$ are 
any
symmetric $\left({}^{0}_{2}\right)$ tensors we have
$$
B(\mu_{\gamma}^{*} h,\mu_{\gamma}^{*} k) = B(h,k),\quad \forall 
\gamma\in G_{x}.
$$
 Because we are assuming the group action 
leaves a metric invariant, such quadratic forms always exist. For example, 
if $g\in \Qhat$, 
then we can set
$$
B(h,k) = g^{\alpha\beta}g^{\gamma\delta} h_{\alpha\gamma}k_{\beta\delta}.
\tag fiber_metric
$$
 Given $h\in 
\sgx$ we can use $B$ to define $\alpha\in \sgstarx$ by 
``raising indices''  in the usual way. 
Thus $B$ defines an isomorphism:
$$
B\colon \sgx\to \sgstarx.
$$
Likewise, it is straightforward to see that, in addition, $B$  
defines an isomorphism:
$$
B\colon S_x^{0} \to \sgperp,
$$
where $\sgperp$ is the orthogonal complement to 
$\sgx$ in $S_x$ with respect to $B$. We thus get an 
isomorphism
$$
B\colon \sgstarx\cap S_x^{0} \to \sgx\cap 
\sgperp,
$$
and the Palais condition is equivalent to
$$
 \sgx\cap 
\sgperp = 0, \quad \forall x\in M
$$
for all $G_{x}$-invariant metrics on the space $S_x$.
Evidently, the Palais condition fails if and only if $\sgx$ and 
its orthogonal complement with respect to every $G_{x}$-invariant 
metric on $S_x$ have common elements. This 
occurs precisely when the restriction of  $B$ to $\sgx$ is degenerate. 
\square

\proclaim Corollary 5.8. 
Condition  \(intersection) is satisfied if there there exists at least 
one positive definite $G_{x}$-invariant scalar product on $T_{x}M$.  

\proof
Using the positive definite scalar product in \(fiber_metric)  the 
resulting metric on $S_x$ is positive-definite and 
will not become degenerate on any subspace. From Lemma 5.7, \(intersection) 
must be satisfied. \square

\proclaim Corollary 5.9.
Condition  \(intersection) is satisfied if there exists a $G$-invariant 
Riemannian metric
on $M$.

\proof
At each point $x\in M$ a $G$-invariant Riemannian metric defines 
a positive definite $G_{x}$-invariant scalar product on $T_{x}M$.
\square

\noindent
This last result shows that PSC2 holds when 
considering symmetry reduction of a theory involving a Riemannian 
metric, that is, in ``Euclidean gravity'' theories.

\proclaim Corollary 5.10.
Condition  \(intersection) is satisfied for group actions with compact 
isotropy groups.

\proof
A compact 
isotropy group will admit a $G_{x}$-invariant positive-definite 
quadratic form at each 
$x$. Once again this leads to a positive definite metric 
\(fiber_metric). \square

\noindent
Note, however, that a compact isotropy group need not 
prevent the failure PSC1. Indeed, this condition can 
fail when the group action is free and the 
group is not unimodular (see below).

 From the preceding considerations 
we obtain a very useful sufficient condition for the 
validity of PSC. 

\proclaim Proposition 5.11.
PSC is valid when $G$ is a compact Lie group.

\proof
As noted in \refto{Anderson1997} 
when $G$ is compact the Lie algebra condition 
\(Lie) is satisfied. 
And, because the isotropy 
subgroups will be compact,  the Palais condition is satisfied
by Corollary 5.10.  \square

This result has been established by Palais 
\refto{Palais1979} for 
PSC defined in terms of
$G$-invariant functions on Banach manifolds.
Note, in particular, that Proposition 5.11 shows that
 PSC is valid when considering 
reductions according to spherical symmetry.

For the rest of this section we specialize to 
the common situation where the isotropy 
subgroups $G_{x}$ are connected so that infinitesimal methods can be 
used to 
characterize the two conditions for PSC. 
  We begin with some definitions. At 
a given point $x\in M$, let $\h\subset\Gamma$ be 
the Lie algebra of the isotropy group $G_{x}$. Define $\m$ to be the 
vector space
$$
\m = \Gamma/\h.
$$
The tangent space to the $G$-orbit at $x$  
can be identified 
with $\m$ and the infinitesimal linear isotropy representation on 
the tangent space to the orbit at $x$ is then identified 
with the adjoint representation $ad_{\h}\colon \h\to gl(\m)$ of $\h$ on 
$\m$:
$$
ad_{{\bf h}} v = [v,{\bf h}]\ {\rm mod}\ \h,\quad v\in \m.
$$
 
The {\it normalizer}   $\n(\h)$ is the largest 
subalgebra of $\Gamma$ that contains $\h$ as an ideal:
$$
[\n(\h),\h]\subset \h.
$$
Let $ad_{\n}\colon {\n}\to gl(\m)$ be the 
adjoint representation of $\n$ on $\m$ given by
$$
ad_{{\bf n}} v = [v,{\bf n}]\ {\rm mod}\ \h,\quad v\in \m.
$$
We define the Lie algebra ${\bf s}\subset\m$ by
$$
{\bf s} = \n(\h)/\h.
$$
It is not hard to see that the isotropy subalgebra 
$\h$ acts trivially on ${\bf s}$.
The space {\bf s} corresponds to the values at $x$ of $G$-invariant vector 
fields tangent to the orbit through $x$. 
We denote by $ad_{\bf s}\colon {\bf s}\to gl(\m)$ the restriction of 
$ad_{\n}$ to $\bf s$.

\proclaim Proposition 5.12. For connected $G_{x}$ with  
Lie algebra $\h$ and normalizer $\n(\h)$,  
\(Lie)  is equivalent 
to 
$$
{\rm tr} (ad_{\n}) = 0,
\tag trad
$$
or equivalently,
$$
{\rm tr}(ad_{\bf h}) = 0,
\tag adh1
$$
$$
{\rm tr}(ad_{\bf s}) = 0,
\tag ads1
$$
for all $x\in M$.

\proof
Pick a complement for $\h\subset\Gamma$ and identify it with $\m$. 
Let us choose a basis $\tau_{i}$ for $\h$ 
and $\tau_{\alpha}$ for $\m$.
In terms of structure constants, condition \(trad) is 
equivalent to the two conditions:  
\itemitem{(i)}
$$
C_{i \alpha}{}^{\alpha} = 0. \tag af1
$$
\itemitem{(ii)} If $v^{\alpha}\in \m$ satisfies
$$
v^{\alpha}C_{i \alpha}{}^{\beta} = 0, 
\tag af2a
$$
then it also satisfies
$$
v^{\alpha}C_{\alpha\beta}{}^{
\beta} = 0. 
\tag af2b
$$

\noindent
Eq. \(af1) is equivalent to \(adh1) and eq. 
\(af2a)-\(af2b) is equivalent to \(ads1).
As shown in \refto{Anderson1997}, conditions (i) and (ii) are 
equivalent to \(Lie). \square

We remark that the conditions \(trad), \(adh1) and \(ads1) 
 mean that the 
 adjoint actions of $\n$, $\bf h$ and $\bf s$  
on $\m$ are each the infinitesimal action of a unimodular group.
In addition,  $v\in {\bf s}$ if and only if it satisfies \(af2a), and 
then $v$ represents the value 
of a $G$-invariant vector field $V$ tangent to the orbit at $x$. 
Condition \(af2b) is  equivalent to the statement that the vector 
field $V$ is divergence-free (at $x$) relative to any $G$-invariant volume 
element on the orbit (\cf 
Proposition 5.4).

\proclaim Proposition 5.13.
If $G_{x}$ is connected and 
$\m$ admits a non-degenerate bilinear form invariant under the 
action of $\h$, \(Lie) is equivalent to
$$
{\rm tr}(ad_{\bf s}) = 0
\tag ads
$$
at all $x\in M$.

\proof
As in Proposition 5.12, pick a complement for 
$\h\subset\Gamma$ and identify it with $\m$.
Denote by $\tau_{i}$ a basis  for $\h$ 
and denote by $\tau_{\alpha}$  a basis for $\m$.
Let $\sigma_{\alpha\beta}$ be the components of the quadratic form
$\sigma\colon \m\times\m\to \reals$ in the chosen basis for \m.
By (infinitesimal) $G_{x}$ invariance, these 
components satisfy
$$
C_{i\alpha}{}^{\gamma}\sigma_{\gamma\beta} + 
C_{i\beta}{}^{\gamma}\sigma_{\alpha\gamma} = 0,\quad\forall\ i.
$$
Contraction with $\sigma^{\alpha\beta}$ 
implies that \(adh1) (equivalently, \(af1)) is satisfied in Proposition 5.12. 
\square

If there exists a $G$-invariant metric on 
spacetime that induces a non-degenerate metric on the orbits, so that 
the orbits are either spacelike or timelike in the Lorentzian case, 
then  this defines a non-degenerate bilinear form on $\m$. We 
therefore have the following result.

\proclaim Corollary 5.14. For a group action with $G_{x}$ connected such that 
there exists a $G$-invariant metric with respect to which the 
group orbits  are non-null, the Lie algebra condition \(Lie) is satisfied  if and only if 
$$
{\rm tr}(ad_{\bf s}) = 0
\tag new_norm
$$
for all $x\in M$.

We remark that 
the Lie algebra condition \(Lie) will be satisfied according to Proposition 
5.13 or 5.14 whenever $\bf s$ is trivial or its representation on $\m$ is 
trivial. The former case occurs when there are no $G$-invariant 
vector fields tangent to the orbits so that ${\bf s}=0$. 
The latter case occurs when the 
$G$-invariant vector fields tangent to the orbits correspond to elements 
of the 
center {\bf c} of the Lie algebra, \ie
 $\n(\h) = \h +{\bf k}$, where ${\bf k}\subset {\bf c}$.  These two situations 
occur frequently among the vector field algebras  listed in \refto{Petrov1969}.  

Let us briefly consider the special case where the group action is 
free, that is, the isotropy group at each point is trivial.  
This case is already treated in \refto{Anderson1997} (see also 
\refto{Shepley1998}); here it becomes a 
corollary to Proposition 5.12. We define $ad_{\Gamma}$ to be the 
infinitesimal adjoint representation of the Lie algebra $\Gamma$ on 
itself.

\proclaim Corollary 5.15. For a free, connected group action PSC is equivalent 
to the unimodular condition
$$
{\rm tr}(ad_{\Gamma}) =0.
\tag freeunimodular
$$

\proof
For a free action
$\h=0$ and $\m={\bf s}=\n=\Gamma$. 
The condition \(trad) becomes simply \(freeunimodular). 
 Because the group action is free, 
the Palais condition \(intersection) for PSC is satisfied. Thus \(freeunimodular) is 
necessary and sufficient for PSC when the group $G$ is connected and 
acting freely.
\square

When the symmetry
group is acting freely with spacelike hypersurface orbits 
then we are considering  spatially 
homogeneous cosmological models. In a four-dimensional spacetime these 
models are often called the ``Bianchi models'' since they are 
determined by a choice of connected three-dimensional Lie group, all 
of which have been classified by Bianchi. The condition 
\(freeunimodular), which is equivalent to the structure constant 
condition
$$
C_{ij}{}^{j}=0,
$$
picks out the Bianchi class A cosmological 
models as obeying 
PSC \refto{MacCallum1972,Sneddon1976,Shepley1998}.

Let us conclude this section by  giving a simple necessary and 
sufficient condition for the Palais condition under the hypothesis that the 
isotropy group of  
any given point is 
connected. 
\proclaim Proposition 5.16. 
Let $G$ act by isometries on a four dimensional Lorentz manifold 
with $G_{x}$ connected for all $x\in M$.  A  necessary and sufficient 
condition for \(intersection) is that the linear 
isotropy representation at any $x\in M$ not correspond to one of the the null 
rotation subgroups. 

\proof
Since at each point $x\in M$ the linear isotropy 
representation of $G_{x}$ must preserve a quadratic form of Lorentz 
signature, it follows that the linear isotropy representation defines 
a conjugacy class of a subgroup of the Lorentz group. Granted that $G_{x}$ is 
connected, this conjugacy class is completely characterized by its 
Lie algebra. Up to conjugation, there are 14 distinct subalgebras of 
the Lorentz Lie algebra ranging in dimensions from 1 to 6 \refto{Winternitz1975}.
It is straightforward to 
compute  the intersection appearing in 
\(intersection) for each of the 14 subalgebras of the Lorentz algebra. 
We find that the
subalgebras generating {\it null rotation} subgroups are the only 
subalgebras that correspond to
infinitesimal linear 
isotropy representations  
which fail to satisfy \(intersection). A null rotation subgroup
of the Lorentz group has the property that it leaves invariant one 
and only one null vector $N$. 
The tensor $N\otimes N$  defines a non-vanishing 
element of $\sgstarx\cap \sgxo$ (there may be other 
non-vanishing elements). \square

There are 3 null rotation conjugacy classes of $SO(3,1)$ with
dimensions 1, 2, and 3. Given an isometry group, 
there is a straightforward, coordinate/frame independent way  of 
testing whether the linear isotropy representation at a point 
corresponds to one of the null rotation subgroups thus violating PSC.  
First, at a given point $x\in M$, one computes the infinitesimal 
linear isotropy 
representation as a Lie algebra of linear transformations of $T_{x}M$. We 
assume that the linear isotropy representation is of dimension 1, 2 or 
3, otherwise \(intersection) is satisfied. 
If the dimension is 1 or 2, 
then the linear isotropy representation corresponds to a null rotation 
if and only if each element of the infinitesimal linear isotropy algebra has 
vanishing eigenvalues. If the dimension of $G_{x}$ is 3, then $G_x$ corresponds
to a null 
rotation if the infinitesimal linear isotropy algebra
admits an element with 
2 non-zero imaginary eigenvalues and an element admitting only zero 
eigenvalues.

\subhead{5.4 Summary of key results on PSC}

Let us  summarize the salient results of this section. Let a group $G$ act 
on $M$ with orbits of fixed dimension $l$ and such that 
the regularity assumptions  given in \refto{CGT2000, Anderson2001} are satisfied. 
Denote the Lie algebra of $G$ by $\Gamma$ and the isotropy group of 
$x\in M$ by $G_{x}$. Denote by $\sgx$ the vector space of 
$G_{x}$-invariant symmetric $\left({}^{0}_{2}\right)$ tensors at 
$x$,  denote by 
$\sgstarx$ 
the vector space of 
$G_{x}$-invariant symmetric $\left({}^{2}_{0}\right)$ tensors at $x$, 
and denote by $\sgxo\subset S_x^{*}$ the annihilator 
of $\sgx$. Let ${\cal H}^{l}(\Gamma,G_{x})$ be the  Lie 
algebra cohomology at degree $l$ of $\Gamma$ relative to $G_{x}\subset 
G$.

\proclaim Theorem 5.17.
The following two conditions are necessary and sufficient for
PSC:
\itemitem{(i)} ${\cal H}^{l}(\Gamma,G_{x}) \neq 0, \ \forall\ x\in M$\quad 
(Lie algebra condition)
\itemitem{(ii)} $\sgstarx\cap \sgxo = 0,\ \forall\ x\in M.$\quad (Palais condition)
\smallskip
In addition, 
\itemitem{$\bullet$} if $G$ is compact then PSC is valid;
\smallskip
\itemitem{$\bullet$} if the group action is free then PSC is valid if 
and only if
$G$ is unimodular;
\smallskip
\itemitem{$\bullet$} if there exists a $G$-invariant Riemannian metric
on $M$,
then PSC is valid if and only if the Lie algebra condition holds;
\smallskip
\itemitem{$\bullet$} in four dimensions, if the metrics under consideration are 
Lorentzian and $G_{x}$ is connected, 
PSC is satisfied if and 
only if the Lie algebra condition holds and $G_{x}$ is not equivalent to a 
null rotation subgroup of the 
Lorentz group, $\forall x\in 
M$. 

All these results on PSC have local, infinitesimal 
versions, which are obtained by replacing the group $G$ acting on $M$ with 
a Lie algebra of vector fields $\Gamma$ on an open set $U\subset M$. In 
particular, in Theorem 5.17 we replace $G_{x}$ with $\Gamma_{x}$ in 
order to get necessary and sufficient conditions for PSC in the 
infinitesimal setting. Of course, 
when 
$\Gamma$ and $\Gamma_{x}$ are the 
infinitesimal generators of a connected $G$ 
action with connected isotropy $G_{x}$ at each $x$, then the 
validity of PSC using $\Gamma$ 
and $\Gamma_{x}$ is equivalent to the validity of PSC using $G$ and $G_{x}$.

\taghead{6.}
\head{6. Examples}

We begin 
by briefly considering  examples in which the isotropy groups are 
 zero-dimensional, \ie trivial or discrete.  We next 
examine in some detail various examples involving four-dimensional group 
actions with 
three dimensional orbits. These examples are 
particularly nice since (i) the three dimensional orbits force the 
reduced field equations to be ODEs, and (ii) the one-dimensional 
isotropy group at each point reduces the number of arbitrary functions 
in the metric to 4, so that the various formulas for the field equations, 
Lagrangians, and Euler-Lagrange equations are manageable. Finally we 
consider a couple of transitive group actions. One novel feature of this latter 
class of examples is that the field 
equations and Euler-Lagrange equations are always algebraic in the 
free parameters that characterize the group-invariant metrics.

In each of the following examples we define 
the (local) symmetry group action, either explicitly or infinitesimally, and 
we check the validity of PSC according to the results of \S 5. We also 
compute the invariant metrics, the reduced Einstein equations, the 
reduced Einstein-Hilbert Lagrangian, and the Euler-Lagrange 
equations of the reduced Lagrangian so as to illustrate the validity or 
failure of PSC.  In each of the examples the  invariant metrics  
on $M$ can be expressed as
$$
g(q) = q^{i}h_{i},
$$
where the $q^{i}\colon M\to \reals^{m}$ are $G$- invariant functions on $M$
and the 
$h_{i}$ are a basis
for $\sg$. 
Since $G$ invariant functions on $M$ are in 1-1 correspondence with 
functions on $M/G$, we can identify the functions $q^{i}$ with 
the fields $\hat q$. This identification
$
\hat q \leftrightarrow q
$
allows us to perform computations on the reduced spacetime manifold by 
using invariant 
functions on $M$. In particular, the field equations, when 
restricted to the $G$-invariant metrics on $M$, can be expressed as
$$
E(\lambda)(g(q)) = \Delta(g(q))\otimes \epsilon = 0,
$$
with
$$
\Delta(g(q)) = \Delta_{i}(q) f^{i},
$$
where $f^{i}$ are a basis for $(\s^*)^{\ss G}$.
The functions $\Delta_{i}(q)$ are a set of $G$ invariant 
differential operators on $M$. 
The reduced equations $\hat\Delta(\q)=0$ are identified with the 
equations $\Delta_{i}(q)=0$. Likewise, the reduced Lagrangian is 
identified with the invariant form 
$$
\hat\lambda(q)=\chi\hook\lambda(g(q))
$$ 
on 
$M$, from which we can compute representatives 
of the Euler-Lagrange equations of the reduced Lagrangian using the 
identification $E(\hat\lambda(\q))=0\Longleftrightarrow 
E(\hat\lambda(q))=0$.

We remind the reader that all of these considerations apply in the 
infinitesimal setting in which $G$ is replaced with $\Gamma$, $G_{x}$ 
is replaced with $\Gamma_{x}$, and so forth.

\subhead{6.1 Freely acting Abelian groups.}

Free Abelian group actions are found in the 
``one Killing vector models'' and the ``two Killing vector models'', 
\etc 
For simplicity we will focus on the case of a two dimensional group. 
Let us suppose that $M=\reals^{4}$ (although other topologies are 
possible)
with coordinates 
$x^{\alpha}$, $\alpha=1,\ldots,4$. We consider the translation group 
action  whose infinitesimal generators are
$$
X_1 = \partial_{1},\quad X_{2}=\partial_{2}.
\tag 2kv
$$
Points in the reduced spacetime manifold $M/G\approx \reals^{2}$ can be 
labeled by 
$(x^{3},x^{4})$.
Every $G$-invariant metric can be 
expressed as
$$
g(q) = q_{\alpha\beta}(x^{3},x^{4}) dx^{\alpha}\otimes 
dx^{\beta},\quad q_{\alpha\beta}=q_{\beta\alpha}.
$$
Thus the space of $G$-invariant metrics can be viewed as the space of 
fields $q_{\alpha\beta}$ on $M/G$.

Freely acting groups trivially satisfy PSC2.
Moreover, for such group actions PSC1 is satisfied
since  the group is unimodular (see 
Corollary 5.15). Indeed, for free Abelian group actions (1) all 
left-invariant forms on $G$ are 
right-$G_{x}$-basic since $G_{x}$ is trivial, and (2) 
all left $G$-invariant forms on the 
group are closed -- so there cannot be any exact forms, and every form 
represents a cohomology class. Thus we have
$$
{\cal H}^{2}(\Gamma,Id) =\reals.
$$
and
a cochain map always exists according to Proposition 5.5. 
 Up to a multiplicative constant, it is 
induced by the 
$G$-invariant chain
$$
\chi=\partial_{1}\wedge \partial_{2}.
\tag 2kvchain
$$
Using the fact that all $G$-invariant vector fields are of the form
$v^{\alpha}\partial_{\alpha}$ with 
$v^{\alpha}=v^{\alpha}(x^{3},x^{4})$, it is easy to check that $\chi$ 
in \(2kvchain) satisfies Proposition 5.4 
and  defines a cochain map via \(rhodef).

The reduced Einstein equations are rather well studied in this 
context (see, for 
example, \refto{MacCallum1980}), 
and can be obtained by
simply dropping all derivatives with respect to $x^{1}$ and $x^{2}$ in 
the field equations. 
Likewise, the reduced Lagrangian $\hat \lambda$ can be obtained by 
simply dropping 
all derivatives with respect to $x^{1}$ and $x^{2}$ in the 
Einstein-Hilbert Lagrangian and then contracting 
with the chain \(2kvchain). It is  a standard exercise to see that 
the Euler-Lagrange equations of $\hat\lambda$ must agree with the 
reduced field equations. Of course, this result is also guaranteed by 
our general theory, \ie 
the fact that $\chi$ defines a cochain map.

\subhead{6.2 Orthogonally transitive group actions} 

We next consider the special case of 
orthogonally transitive two dimensional Abelian isometry groups. 
Recall that a group action is orthogonally transitive if, for 
any $G$-invariant metric, the 
distribution ${\cal D}\subset TM$ orthogonal to the group orbits is 
integrable. This definition is slightly awkward since it uses the 
infinite dimensional space of $G$-invariant metrics to characterize 
the isometry group action. As mentioned in \S 2, 
we prefer to view the group action as simply a Lie group $G$
of diffeomorphisms of a manifold $M$, which then is used to select 
the allowed $G$-invariant metrics.  For our purposes, a 
superior -- but equivalent -- definition of an orthogonally 
transitive group action is as the semi-direct product of a
discrete ($Z_{2}$) group, with action
$$
(x^{1},x^{2}) \to (-\epsilon x^{1},-\epsilon x^{2}),\quad \epsilon = 
\pm 1,
\tag discrete
$$
and the $(x^{1},x^{2})$ translation group, which  is 
a normal subgroup. The non-trivial 
action of the isotropy group $G_{x_{0}}$ of a point with 
coordinates $x_{0}^{\alpha}$ is given by
$$
x^{\alpha}\to (-x^{1} + 2x^{1}_{0},-x^{2} + 2x_{0}^{2},x^{3},x^{4}).
$$
The reduced spacetime manifold $M/G$ can still be parametrized by 
$(x^{3},x^{4})$.
Because of the isotropy constraint \(iso_cond) the
$G$-invariant metrics are now of the form
$$
g(q) = q_{ab}(x^{i}) dx^{a}\otimes dx^{b} + 
\tilde q_{ij}(x^{i})dx^{i}\otimes dx^{j},\quad q_{ab}=q_{ba},\ 
q_{ij}=q_{ji}
\quad
a,b=1,2\quad i,j=3,4,
\tag 2kvmetric
$$
so that $\Qhat$ is the set of fields parametrized by $q_{ab}$ and 
$\tilde q_{ij}$ on 
$M/G$, constrained so that $g(q)$ has Lorentz signature. 

From \(2kvmetric) it is clear that, with respect to any $G$-invariant metric, 
the orbit-orthogonal distribution $\cal D$ is integrable. Conversely, 
according to the traditional definition, every metric admitting an 
orthogonally transitive group action can be put into the form 
\(2kvmetric), so that 
the group action and $G$-invariant metrics we have described are equivalent 
to the usual notion of an 
orthogonally transitive group action.

Let us now consider the two conditions for PSC. Of course, it is well known 
that the Einstein-Hilbert Lagrangian does reduce to give a valid 
Lagrangian for the $G$-invariant spacetimes (see, \eg \refto{CGT1996}). 
We shall see 
that, in fact, there 
are no obstructions to PSC, thus guaranteeing the successful 
reduction of {\it any} ($G$-invariant) Lagrangian. 

The Lie algebra condition for PSC can be computed as follows. Fix 
a basis $(e_{1},e_{2})$ for the Abelian Lie algebra $\Gamma$ of the 
isometry group. 
The 
dual basis of left-$G$-invariant forms on $G$ are denoted 
$(\omega^{1},\omega^{2})$ and satisfy
$$
d\omega^{1} = 0 = d\omega^{2}.
$$
The right-action of $G_{x}$ on the dual basis is given by
$$
\omega^{i}\longrightarrow-\omega^{i},\quad i=1,2.
$$
The closed, right-$G_{x}$-basic 2-forms on $G$ are proportional to
$$
\omega = \omega^{1}\wedge\omega^{2}.
$$
Since all left-$G$-invariant forms are closed, $\omega$ cannot be 
exact and we see that ${\cal H}^{2}(\Gamma,G_{x})=\reals$ for all 
$x\in M$.  Equivalently, it is easy 
to see that the chain from the previous example,
$$
\chi = \partial_{1}\wedge\partial_{2},
$$
is $G$-invariant and still defines a cochain map.

To check the Palais condition \(intersection) for PSC we compute
$$
\sgx = \{dx^{a}\odot dx^{b}, dx^{i}\odot dx^{j}\},\quad 
a,b=1,2\quad i,j=3,4,
$$
$$
\sgstarx = \{\partial_{a}\odot \partial_{b}, \partial_{i}\odot 
\partial_{j}\},\quad 
a,b=1,2\quad i,j=3,4,
$$
and
$$
\sgxo = \{\partial_{a}\odot \partial_{i}\}
$$
so that
$$
\sgstarx\cap \sgxo = 0.
$$
(Here we introduce the notation in 
which $\{W\}$ is the vector space spanned by the vectors $W$).

These computations show that PSC is satisfied for the orthogonally 
transitive 2 Killing vector models, irrespective of the choice of 
Lagrangian for the metric. This is, of course, easy to verify 
directly. For example, consider the vacuum Einstein theory described 
by the Einstein-Hilbert Lagrangian. 
The reduced Lagrangian is obtained  as in the free group case 
considered previously by (1) dropping all derivatives with respect to
$x^{1}$ and $x^2$, (2) setting $g_{ai}=0$, (3) contracting with the 
chain $\chi$. It then follows that the 
Euler-Lagrange equations of the reduced Lagrangian correspond to the 
field equations 
$$
\G_{ab}=0=\G_{ij}
\tag oteqns
$$
for the metrics of the form \(2kvmetric). 
It is easy to check that for the metrics \(2kvmetric) the 
reduced field 
equations are equivalent to \(oteqns). In particular, the missing 
components $\G_{ai}$ are identically zero for metrics \(2kvmetric) 
admitting the orthogonally transitive group action. This follows from 
the $G$-invariance of the Einstein 
tensor when evaluated on a $G$-invariant metric.

\subhead{6.3 Stationary, spherically symmetric spacetimes}

Here we consider the well-known case of a spacetime admitting time 
translational
and rotational  symmetry.  Of course, for reductions by the group $SO(3)$ PSC 
is guaranteed thanks to the compactness of 
$SO(3)$ (see Proposition 5.11). But the addition of time translation symmetry 
renders the 
symmetry group non-compact, so a closer look is warranted.

The spacetime manifold can be taken to be $M= \reals^{+}\times \reals\times
{S^{2}}$.  The abstract Lie group is 
$G= \reals\times SO(3)$ 
with orbits $ \reals\times S^{2}$. The isotropy group $G_x$ of any point is 
isomorphic to $SO(2)\subset SO(3)$.
The reduced spacetime is $M/G\approx \reals^{+}$. Since  $G$ and 
$G_{x}$ are connected, 
we  will use infinitesimal methods to study PSC.

The $G$-invariant metrics are
$$
g(q) =   q^{1}(r) dt\otimes dt + q^{2}(r) dt\odot dr
+q^{3}(r) dr\otimes dr +\half q^{4}(r) d\Omega^2  = q^{i}h_{i},
\tag SO3_metric
$$
where  $r$ labels the group orbits, $dt$ and $dr$ are $G$-invariant 1-forms, 
$d\Omega^2$ is the standard metric on $S^2$,
$q^{i}=q^{i}(r),\  i=1,2,3,4$ with
$$
(q^{4})^{2}\left((q^{2})^{2} - 4q^{1}q^{3}\right) >0
$$ 
 are the dependent variables for the 
 reduced theory, 
and the 
$h_{i}$ are a basis for the $G$-invariant symmetric 
$\left({}^{0}_{2}\right)$ tensor fields. 

About any given point in $M$  we will use a spherical polar coordinate chart 
$(t,r,\theta,\phi)$, where 
$$
d\Omega^2=(d\theta\otimes d\theta + 
\sin^{2}\theta\,  d\phi\otimes d\phi).
$$
Points in $M/G$ 
will be labeled by  $r$. 
 A set of infinitesimal generators of the group action are expressed 
in the coordinate chart as
$$
X_{1} = {\partial\over\partial\phi},\quad
X_{2}=\sin\phi{\partial\over\partial\theta} + 
\cos\phi\cot\theta{\partial\over\partial\phi},
\quad
X_{3}=\cos\phi{\partial\over\partial\theta} - 
\sin\phi\cot\theta{\partial\over\partial\phi},
\quad
X_{4} = {\partial\over\partial t}.
$$ 
This algebra of vector fields $\Gamma$ spanned by 
$(X_{1},\ldots,X_{4})$ appears in Petrov's classification 
\refto{Petrov1969} as (32.9). 
At a generic 
point, $x_{0}^{\alpha}=(t_{0},r_{0},\theta_{0},\phi_{0})$ 
the infinitesimal generator 
of $G_{x_{0}}$
is given in coordinates by
$$
Y=-\cot\theta_{0} X_{1} + \cos\phi_{0} X_{2} - \sin\phi_{0} X_{3}.
$$
The infinitesimal linear isotropy representation at $x_{0}$ is then given in this chart 
by
$$
Y^{\alpha}_{,\beta}\Big|_{x_{0}} = \left(
\matrix{0&0&0&0\cr
0&0&0&0\cr
0&0&0&1\cr
0&0&-{\rm csc}^{2}\theta_{0}&0}\right).
\tag lir329
$$

Let us first consider PSC2.
The eigenvalues of the matrix \(lir329) constitute a complex 
conjugate, pure imaginary pair corresponding to the fact that 
$G_x=SO(2)$.  We conclude, by Proposition 5.16, 
that the Palais condition -- and hence PSC2 -- is satisfied.

We next consider PSC1. The Lie group $G$ and its isotropy subgroups are 
connected and the orbits of the group 
action are 
reductive homogeneous spaces ${\reals\times SO(3)\over SO(2)}$. 
Let us view $\Gamma$ as an abstract Lie algebra with basis 
$X_{i}$, $i=1,2,\ldots,4$. At the point $x_{0}$, defined by 
$t=const$, $r=const.$, $\theta={\pi\over 2}$, 
$\phi={\pi\over 2}$, we have the reductive decomposition $\Gamma=\h+\m$, where
$$
{\bf h} = \{X_{3}\},\quad {\bf m}=\{X_{1},X_{2},X_{4}\}.
$$
It is easy to check that, at any point,
$$
{\bf n}({\bf h})\cap {\bf m} = \{X_{4}\},
$$
and that $X_{4}$ spans the center of $\Gamma$. Moreover, 
the $G$-invariant metrics include the Minkowski 
metric, which induces a non-degenerate metric on the group orbits. 
According to 
Corollary 5.14 the Lie algebra condition -- and hence PSC1 --  is satisfied. 

It is straightforward to compute ${\cal H}^3(\Gamma,G_x)$ directly.
 Denote by $\omega^{i}$, $i=1,2,3,4$ the 
basis of left-invariant differential forms on 
$G$ dual to the basis $X_{i}$. This dual basis satisfies
$$
d\omega^{i}=-\half\epsilon^{i}{}_{jk}\omega^{j}\wedge \omega^{k},\quad 
i,j,k=1,2,3,
$$
and
$$
d\omega^{4}=0.
$$
At the point $x_{0}$ defined above, the right-$G_{x_{0}}$-basic closed $3$-forms 
are proportional to 
$\omega^{1}\wedge\omega^{2}\wedge\omega^{4}$ while 
the right-$G_{x_{0}}$-basic 2-forms are all proportional to 
$\omega^{1}\wedge \omega^{2}$. 
Since
$$
d(\omega^{1}\wedge \omega^{2}) = 0,
$$
it follows that $\omega^{1}\wedge\omega^{2}\wedge\omega^{4}$ cannot 
be exact, and hence ${\cal H}^{3}(\Gamma,G_{x_{0}})=\reals$. It is not 
hard to see that this result is valid for any $x\in M$.

From our general theory we know that a suitable cochain map exists. 
In a spherical polar coordinate chart
it is given by
$$
\chi = {1\over\sin \theta} {\partial \over \partial 
t}\wedge{\partial\over\partial \theta}\wedge{\partial\over\partial 
\phi}.
\tag SO3_cochain
$$
$\chi$ is $G$-invariant and
satisfies
$$
L_{\ss \partial\over\partial t} \chi = 0 = L_{\ss \partial\over\partial r}\chi.
$$
Note that any invariant vector field is a combination of
$(\partial_{t},\partial_{r})$ with $r$-dependent coefficients, 
so that  a cochain map 
exists according to Proposition 5.4.

Let us consider the symmetry reduction of the  vacuum Einstein 
equations $\G(g) = 0$. 
The restriction of the Einstein tensor to a $G$-invariant metric $g(q)$ 
is given by
$$
\G(g(q)) = \G_{i}(q) f^{i},
$$
where $\G_{i}$ are second order differential functions of the $q^{i}$, 
and the $f^{i}$ are a basis of $G$-invariant symmetric 
$\left({}^{2}_{0}\right)$ tensors dual to the $h_{i}$. In the spherical 
polar coordinate chart 
they are:
$$
f^{i}=\left({\partial \over\partial t}\otimes {\partial \over\partial t},
{\partial \over\partial t}\odot {\partial \over\partial r},
{\partial \over\partial r}\otimes {\partial \over\partial r},
{\partial \over\partial \theta}\otimes {\partial \over\partial 
\theta}
+{1\over \sin^{2}\theta}{\partial \over\partial \phi}\otimes {\partial 
\over\partial \phi}\right).
\tag f329
$$
With some purely algebraic rearrangements, the 4 field equations 
$\G_{i}=0$ are equivalent to 3 algebraically independent equations
$$
q^{1\prime}=-{q^{1}(q^{4\prime})^{2} + 2q^{4}\over 2q^{4}q^{4\prime}},
\tag SO3_eq1
$$
$$
q^{1\prime\prime} = {q^{1} (q^{4\prime})^{2} + 2q^{4}\over 
2(q^{4})^{2}},
\tag SO3_eq2
$$
$$
q^{4\prime\prime} = {(q^{4\prime})^{2}\over2 q^{4}},
\tag SO3_eq3
$$
where
$$
{F}^{\prime} := {1\over \sqrt{ (q^{2})^2-4q^{1}q^{3}}} {d\over dr} F.
\tag SO3_eq4
$$
The fact that only 3 field equations are algebraically independent is 
a consequence of the contracted Bianchi identities, which provide an 
algebraic relationship among the four field equations that are 
allowed by $G$-invariance. 

Let us now consider the  reduction of the Einstein-Hilbert 
Lagrangian,
$$
\lambda =  R(g)\,\epsilon(g).
$$
Restricting this 
Lagrangian to the $G$-invariant metric $g(q)$ in 
\(SO3_metric) yields
$$
\lambda(g(q))= {1\over2 q^{4}}
\sqrt{(q^{2})^{2}-4q^{1}q^{3}} \left(4 q^{1}q^{4}q^{4\prime\prime} - 
q^{1}(q^{4\prime})^2 + 2 (q^{4})^{2} q^{1\prime\prime} + 
4q^{4}q^{4\prime}q^{1\prime} + 2 q^{4} \right) \nu,
$$
where, in spherical polar coordinates,
$$
\nu=\sin\theta\, dt\wedge dr\wedge d\theta\wedge d\phi.
$$

Using \(SO3_cochain), and using the radius $r$ as a coordinate on 
$M/G$, the reduced Lagrangian is given by
$$
\eqalign{
\hat\lambda(q) &=\chi\hook(\lambda(g(q)))\cr
&= {1\over2 q^{4}}
\sqrt{(q^{2})^{2}-4q^{1}q^{3}} \left(4 q^{1}q^{4}q^{4\prime\prime} - 
q^{1}(q^{4\prime})^2 + 2 (q^{4})^{2} q^{1\prime\prime} + 
4q^{4}q^{4\prime}q^{1\prime} + 2 q^{4} \right) dr.}
$$
A straightforward, if a bit lengthy, computation reveals that the 4
Euler-Lagrange equations of $\hat\lambda(q)$ are equivalent to the 
equations $\G_{i}=0$, or \(SO3_eq1)--\(SO3_eq4). In fact, denoting by $E_{i}$ the 
Euler-Lagrange expressions obtained for each of the $q^{i}$, we have that
$$
E_{i}(\hat\lambda(q))= -q^{4}\sqrt{(q^{2})^{2} - q^{1}q^{3}}\,\,\G_{i}.
$$

\subhead{6.4 Locally isotropic Bianchi class B}

This example can be viewed as a special case of the Bianchi class B 
cosmological models mentioned in the introduction. 
In local coordinates $(\lambda^1,\lambda^2, \lambda^3)\in \reals^3$, 
$\lambda^4\in (0,2\pi)$  on $G$ and in coordinates $x^{\alpha}$,
$\alpha=1,\ldots,4$ on $M=\reals^{4}$
we define a four dimensional group action
$$
x^{\alpha}\longrightarrow \mu^{\alpha}(\lambda,x)
$$
by
$$
\mu^{\alpha}(\lambda,x)
=\left(
 x^{1}-\lambda^{3},
  \lambda^{1}
 +e^{\lambda^{3}}[x^{2} \cos(\lambda^{4})-x^{3} \sin(\lambda^{4})],
       \lambda^{2} + e^{\lambda^{3}}[x^{3} \cos(\lambda^{4}) 
       + x^{2} \sin(\lambda^{4})], x^{4}\right).
\tag 326group
$$
The reduced 
spacetime $M/G\approx \reals$ can be parametrized by $x^{4}$. 
The isotropy groups $G_{x}$ are all connected and isomorphic to $SO(2)$. For 
example, at the origin $x_{0}^{\alpha}=(0,0,0,0)$ the action of $G_{x_0}$ is 
given in local coordinates by
$$
\mu^\alpha(\lambda,x) = (x^{1},x^{2} \cos(\lambda^{4})-x^{3} \sin(\lambda^{4}),
x^{3} \cos(\lambda^{4}) 
+ x^{2} \sin(\lambda^{4}), x^{4}).
\tag 326iso
$$
The group action \(326group) has a Lie algebra $\Gamma$ of infinitesimal generators 
spanned by
$$
X_{1}={\partial\over\partial x^{2}},\quad
X_{2}={\partial\over\partial x^{3}},\quad
X_{3}=-{\partial\over\partial x^{1}}+x^{2}{\partial\over\partial 
x^{2}} +x^{3}{\partial\over\partial 
x^{3}},\quad
X_{4}=x^{2}{\partial\over \partial x^{3}}-x^{3}{\partial\over \partial x^{2}},
\tag gamma326
$$
which 
corresponds to (32.6) in \refto{Petrov1969}. 
The generator of the isotropy subgroup of the origin, \(326iso),  is $X_{4}$.

The 
general form of a $G$-invariant metric is
$$
g(q)=\half q^{1}e^{2x^{1}}\left[dx^{2}\otimes dx^{2}+   
dx^{3}\otimes dx^{3}\right] +q^{2} dx^{1}\otimes dx^{1} 
- q^{3} dx^{1}\odot dx^{4} + q^{4} dx^{4}\otimes dx^{4}
=q^{i}h_{i},
\tag metric32.6
$$
where $q^{i}=q^{i}(x^{4})$, $i=1,\ldots,4$ are arbitrary except for 
the signature requirement
$$
(q^1)^{2}\left[4\,q^2\,q^4-(q^3)^2\right]<0.
$$

We first consider the Lie algebra condition for PSC. Because $G$ and $G_{x}$ 
are connected, we can use Proposition 5.12 to check \(Lie). 
Let us view $\Gamma$ as an abstract Lie algebra with 
basis $X_{i}$, $i=1,2,\ldots,4$.  At $x_{0}^{\alpha}=(0,0,0,0)$ the 
isotropy subalgebra is generated by $X_{4}$ and we have the reductive 
decomposition $\Gamma=\h+\m$, with ${\bf h}=\{X_{4}\}$ and 
${\bf m}=\{X_{1},X_{2},X_{3}\}$. In addition, we have that (at $x_{0}$) 
$\n(\h)\cap \m=\{X_{3}\}$. Since
$$
\sum_{j=1}^{3} C_{3}{}_{j}^{j} = -2,
$$
we see that \(ads1)
is not satisfied. It is not hard to see that a similar result occurs 
at each point in $M$, and so PSC1 fails. 

Let us check the failure of PSC1 
by explicitly computing ${\cal H}^3(\Gamma,G_{x_0})$. Using the basis 
$\omega^{i}$ of left-invariant 1-forms on $G$ dual to the 
$X_{i}$,  it is not hard to 
check that all $G_{x_{0}}$ basic closed 3-forms on $G$ are 
proportional to 
$$
\alpha=\omega^{1}\wedge\omega^{2}\wedge\omega^{3}.
$$
It is likewise straightforward to check that
$$
\beta = \omega^{2}\wedge\omega^{1}
$$
is a $G_{x_{0}}$ basic 2-form and that
$$
\alpha = d\beta,
$$
so that
$$
{\cal H}^{3}(\Gamma,G_{x_{0}}) = 0,
$$
confirming the failure of PSC1.

The preceding computations indicate that no suitable 
cochain map will exist. Indeed, it is easy to check that any 
$G$-invariant chain is a function of $x^{4}$ multiplied by 
$$
\chi = e^{-2x^{1}}{\partial\over\partial x^{1}}\wedge{\partial\over\partial x^{2}}
\wedge{\partial\over\partial x^{3}}.
\tag 32.6chain
$$
Any $G$-invariant vector field tangent to an orbit is a function of 
$x^{4}$ multiplied by
$$
S= {\partial \over \partial x^{1}}.
$$
A $G$-invariant chain cannot be invariant under the flow generated by 
a $G$-invariant vector field tangent to the orbits since
$$
L_{\ss S}\chi = -2\chi.
$$
Thus, by Proposition 5.4 there can be no cochain map, \ie PSC1 fails.

As for PSC2, it is straightforward to check that at 
a generic point $x_{0}^{\alpha}$ the isotropy algebra $\Gamma_{x_{0}}$ is 
generated by
$$
Y=x^{3}_{0} X_{1} - x^{2}_{0} X_{2} + X_{4},
$$
so that the infinitesimal linear isotropy representation is given by
$$
Y^{\alpha}_{,\beta}\Big|_{x_{0}} = \left(\matrix{0&0&0&0\cr 
0&0&-1&0\cr
0&1&0&0\cr
0&0&0&0}\right),
$$
which has pure imaginary, complex conjugate eigenvalues.
As in the previous example, the 
linear isotropy representation is that of a spatial rotation, 
$SO(2)$, and the Palais condition for PSC is satisfied.

The restriction of the Einstein tensor $\G$ to a $G$-invariant metric  
is given by
$$
\G(g(q)) := \G_{i}(q) f^{i},
$$
where $\G_{i}$ are second order differential functions of the $q^{i}$, 
and the $f^{i}$ are a basis for $(\s^*)^{\ss G}$. With some purely 
algebraic rearrangements, the equations 
$\G_{i}=0$ are equivalent to
$$
(Dq^{1})^{2} -  {1\over (q^{2})^{2}} = 0,
\tag eq326.1
$$
$$
(q^{2\prime})^{2} - 1 = 0,
\tag eq326.2
$$
$$
(Dq^{1})^{\prime} +{1\over (q^{2})^2}=0,
\tag eq326.3
$$
$$
q^{2\prime\prime} = 0,
\tag eq326.4
$$
where
$$
Dq^{1} = {1\over \sqrt{(q^3)^2-4q^2q^4}}\left({1\over q^{1}}{dq^{1}\over dx^{4}} + 
{q^3\over q^{2}}\right)
$$
and the prime denotes
$$
F^\prime := {1\over \sqrt{(q^3)^2-4q^2q^4}}{dF\over dx^{4}}.
$$

 Restricting the 
Einstein-Hilbert 
Lagrangian to a $G$-invariant metric gives
$$
\eqalign{
\lambda(g(q)) = {e^{2x^{1}}\over 2q^{2}}\sqrt{(q^{1})^{2}
[(q^{3})^{2} - 4q^{2}q^{4}]}
\Big[&-3 + 2 q^{2} q^{2\prime\prime} + 3 
(Dq^{1})^{2}(q^{2})^{2}\cr 
&+ 4 (Dq^{1})^{\prime}(q^{2})^{2} + 4 
q^{2} Dq^{1}q^{2\prime}\Big]dx^{1}\wedge dx^{2}\wedge 
dx^{3}\wedge dx^{4}}
$$
To construct a reduced Lagrangian we need to pick a $G$-invariant 
chain $\chi$.  Since $\chi$ can't be further specified by demanding 
it define a cochain map there is some arbitrariness in our choice 
of $\chi$. In particular,  we can always multiply $\chi$ by any 
function of 
$x^{4}$ and keep the chain $G$-invariant. 
The most general reduced Lagrangian is thus obtained by using
\(32.6chain) multiplied by a function $f(x^{4})$. We then have
$$
\hat\lambda ={f(x^{4})\over 2 q^{2}} \sqrt{(q^{1})^{2}[(q^{3})^{2} - 4 q^{2}q^{4}]}
\left(2q^{2} q^{2\prime\prime} + 3(q^{2})^{2} (Dq^{1})^{2}
+ 4 (q^{2})^{2} (Dq^{1})^{\prime} + 4 q^{2 }q^{2\prime} Dq^{1}
-3\right)
dx^{4}
$$
There is no choice of function $f$ such that the Euler-Lagrange 
equations for $\hat \lambda$ are equivalent to the reduced field 
equations \(eq326.1)--\(eq326.4). For example, direct computation shows 
that
the Euler-Lagrange equations of $\hat\lambda$ imply
$$
Dq^{1} = {2\over q^{2}\sqrt{(q^{3})^{2} - 4 q^{2}q^{4}}}
\left({q^{2}\over f}{d f\over d x^{4}} - q^{3}\right),
$$
which is clearly inequivalent to \(eq326.1) for any choice of $f$.

\subhead{6.5 Null rotation isotropy}

Let us now examine an example for which PSC2 fails. 
Our example is taken from \refto{Petrov1969} (32.8). 
On $\reals^{4}$, introduce  coordinates 
$x^{\alpha}=(t,x^{1},x^{2},x^{3})$ and a Lie algebra $\Gamma$ of vector 
fields spanned by
$$
X_{1}= e^{-x^{3}}\left({\partial\over\partial x^{1}} - (x^{2})^{2} 
{\partial\over\partial x^{2}} - 2 x^{2} {\partial\over \partial 
x^{3}}\right),\quad
X_{2}= {\partial\over\partial x^{3}},\quad
X_{3}= e^{x^{3}}{\partial\over \partial x^{2}},\quad
X_{4}= {\partial\over\partial x^{1}}.
\tag gamma328
$$
The 
isotropy subalgebra $\Gamma_{x_{0}}$ of a point 
$x_{0}^{\alpha}=(t_{0},x^{1}_{0},x^{2}_{0},x^{3}_{0})$ is generated by 
the vector field
$$
Y_{x_{0}} = e^{x^{3}_{0}} X_{1} + 2 x_{0}^{2}X_{2} + 
e^{-x^{3}_{0}}(x_{0}^{2})^{2} X_{3} - X_{4}.
\tag iso328
$$
Note that the vector fields \(gamma328) only generate a 
local group action on $\reals^{4}$ 
so 
we work with the infinitesimal version of PSC.

The general form of a $\Gamma$-invariant 
metric is
$$
\eqalign{
g(q)&=4 q^{1} dx^{1}\odot dx^{2} - 4 q^{1} x^{2} dx^{1}\odot dx^{3} + 
4q^{2}dx^{2}\otimes dx^{2} - 8 q^{2} x^{2} dx^{2}\odot dx^{3}  + 2 q^{3} dt 
\odot dx^{2}\cr
&\quad+ (4 q^{2} (x^{2})^{2} - q^{1})dx^{3}\otimes dx^{3} - 2 q^{3} x^{2} dt\odot 
dx^{3} + q^{4 } dt\otimes dt\cr
&=q^{i}h_{i},}
\tag metric32.8
$$
where $q^{i}=q^{i}(t)$, $i=1,\ldots,4$ are arbitrary except for 
the signature requirement
$$
q^{1} q^{4}<0.
$$
We shall restrict attention to the case where
$$
q^{4}>0,
$$
which implies that the group orbits are  timelike.

PSC1 is satisfied in this example. At the origin the 
isotropy $\Gamma_{x_0}$ is generated by the vector field
$$
Y = X_{1} - X_{4}.
$$
We view $\Gamma$ as an abstract Lie algebra with basis $e_{i}$, 
$i=1,2,3,4$, taken from \(gamma328) and with dual basis $\omega^{i}$. 
The space of $\Gamma_{x_0}$-basic closed 3-forms on $\Gamma$ is spanned
 by the 
form
$$
\alpha = \omega^{1}\wedge \omega^{2}\wedge \omega^{3} - 
\omega^{2}\wedge \omega^{3}\wedge \omega^{4}.
$$
The space of $\Gamma_{x_0}$-basic 2-forms on $\Gamma$ is spanned 
by the form
$$
\beta = \omega^{2}\wedge \omega^{3}.
$$
Since $d\beta=0$, it follows that $H^{3}(\Gamma,\Gamma_{x_0})=\reals$. It 
is easy to see that this same result holds for all $x\in M$. 
A $\Gamma$-invariant chain that 
provides a cochain map is given by constant multiples of
$$
\chi = {\partial\over\partial x^{1}}\wedge 
{\partial\over\partial x^{2}}\wedge
{\partial\over\partial x^{3}}.
\tag chi328
$$
Since the module of $\Gamma$-invariant vector fields is generated by 
$({\partial\over\partial t},{\partial\over\partial x^{1}})$ and since
$$
L_{\ss \partial\over\partial t} \chi = 0= L_{\ss \partial\over\partial 
x^{1}}\chi,
$$ 
it follows from Proposition 5.4 that $\chi$ defines the desired cochain map.

To see that PSC2 fails in this example, 
we compute the 
infinitesimal linear isotropy representation at a generic point 
$x_{0}^{\alpha}=(t_{0},x^{1}_{0},x^{2}_{0},x^{3}_{0})$. We find
$$
Y^{\alpha}_{,\beta}\Big|_{x_{0}} = \left(\matrix{0&0&-1&0\cr
0&-2 x_{0}^{2}&2 (x_{0}^{2})^{2}&0\cr
0&-2&2x_{0}^{2}&0\cr
0&0&0&0}\right)
\tag lir328
$$
The eigenvalues of \(lir328) all vanish, so the infinitesimal linear isotropy representation corresponds to the Lorentz subalgebra of null rotations.
 Thus by Theorem 5.17 the Palais condition fails.

The restriction of the Einstein tensor $\G$ to
a $\Gamma$-invariant metric \(metric32.8)
is given by
$$
\G(g(q)) = \G_{i}(q) f^{i},
$$
where $\G_{i}$ are second order differential functions of the $q^{i}$ 
and the $f^{i}$ are  $\Gamma$-invariant tensor fields given by
$$
\eqalign{f^{i}&=
 \big({\partial\over\partial x^{1}}\odot {\partial\over\partial x^{2}}
-  (x^{2})^{2} {\partial\over\partial x^{2}}\otimes{\partial\over\partial 
x^{2}} - 2x^{2} {\partial\over\partial x^{2}}\odot{\partial\over\partial 
x^{3}}-  {\partial\over\partial x^{3}}\otimes{\partial\over\partial 
x^{3}},\cr &{\partial\over\partial x^{1}}\otimes {\partial\over\partial 
x^{1}},
{\partial\over\partial x^{1}}\odot {\partial\over\partial 
t},
{\partial\over\partial t}\otimes {\partial\over\partial t}
\big).}
\tag fi
$$
With some purely 
algebraic rearrangements, the equations 
$\G_{i}=0$ are equivalent to the three equations
$$
(q^{1\prime})^{2} -q^{1} = 0,
\tag eq328.1
$$
$$
q^{1\prime\prime} - \half = 0,
\tag eq328.2
$$
$$
q^{2\prime\prime}={4 q^{2} + q^{1\prime} q^{2\prime}\over q^{1}},
\tag eq328.3
$$
where
the prime is defined by
$$
F^\prime = {1\over \sqrt{q^4}}{dF\over dt}.
$$
There are only three independent Einstein equations because the 
Bianchi identities provide an algebraic identity satisfied by the 
four field equations allowed by $G$-invariance.

The restriction of the Einstein-Hilbert
Lagrangian to the $\Gamma$-invariant metric \(metric32.8) is 
$$
\eqalign{
\lambda(g(q)) = -3\sqrt{-q^{1}q^{4}}(2 q^{1\prime\prime}-1)
\,dx^{1}\wedge dx^{2}\wedge 
dx^{3}\wedge dt}.
$$
To construct a reduced Lagrangian we use the chain $\chi$ given in 
\(chi328). This yields the reduced Lagrangian
$$
\hat\lambda = -3\sqrt{-q^{1}q^{4}}(2 q^{1\prime\prime}-1)\, dt
\tag redlag328
$$
Note that only two of the four dependent variables appear in the 
Lagrangian; there are only two non-trivial Euler-Lagrange equations, and they are 
equivalent to \(eq328.1) and \(eq328.2). The third field equation 
\(eq328.3) does not follow from the reduced Lagrangian \(redlag328).
That there is 
a single missing equation can be understood as follows.
First, we note that
$
(S_x^{*})^{\ss \Gamma_{x}} = \{f^{i}\}_{x}$, so that using \(fi)
we have
$$
(S_x^{*})^{\ss \Gamma_{x}}\cap (S_x^{\ss\Gamma_{x}})^{0}= 
\{{\partial\over\partial x^{1}}\otimes {\partial\over\partial 
x^{1}}, {\partial\over\partial x^{1}}\odot {\partial\over\partial 
t}\}_{\ss x}
\tag fiber328
$$
showing again that the Palais condition fails.
 Since \(fiber328) is two dimensional, there 
are two field equations which will not 
follow from the reduced Lagrangian. As noted earlier, the Bianchi 
identities imply that one of the field equations
is not algebraically independent of the other three. This 
redundant equation happens to be one of the equations not provided by 
the reduced Lagrangian. So, at the end of the day, the reduced 
Lagrangian fails to provide only one of the independent field equations. 
Note that the Bianchi identities stem from full diffeomorphism 
equivariance -- as opposed to $\Gamma$-equivariance -- of 
the Einstein-Hilbert Lagrangian, so the fact that only one rather than 
two field equations are missing is a consequence of a 
special choice of Lagrangian.

\subhead{6.6 Homogeneous spacetimes}

Homogeneous spacetimes (admitting a transitive isometry group) provide 
examples 
where the reduced field equations are purely algebraic. In addition, 
such symmetry reductions  
provide examples in which PSC is valid, in which PSC1 and PSC2 each 
fails separately, and in which they fail 
simultaneously.  Given our previous examples, we will keep the 
presentation brief.

\bigskip
\noindent
{\it Example 6.6a}

Let us begin with an example, taken from \refto{Petrov1969} eq. 
(33.23), 
 in which PSC is valid.  
On $M=\reals^{4}$ with coordinates 
$x^{\alpha}$, $\alpha=1,\ldots,4$, the Lie algebra of vector fields 
$\Gamma$ is spanned by:
$$
X_{1}=\partial_{2},\quad X_{2}=\partial_{3},\quad
X_{3} = -\partial_{1}+x^{3}\partial_{2},
$$
$$
X_{4}=- x^{3} \partial_{1} + 
\half\left((x^{3})^{2} - (x^{1})^{2}\right)\partial_{2} + x^{1}\partial_{3},\quad 
X_{5}=\partial_{4}.
\tag gamma3323
$$
At each point  $x_{0}\in M$  
the isotropy algebra $\Gamma_{x_{0}}$ is spanned by
$$
Y_{x_{0}} = \half\left((x^{3}_{0})^{2} + (x_{0}^{1})^{2}\right)X_{1}
-x^{1}_{0} X_{2} - x^{3}_{0} X_{3} + X_{4}.
\tag iso3323
$$

By computing the flows of the vector fields \(gamma3323) one can 
compute the corresponding connected group action $(\mu,G)$ 
on $\reals^{4}$ with 
connected isotropy group $G_{x_{0}}$ generated by \(iso3323). 
The spacetime manifold is thus identified as the homogeneous space 
$\reals^{4} \approx G/G_{x_{0}}$.  Because both $G$ and $G_{x_{0}}$ 
are connected, the relevant computations for PSC can all be performed 
using its infinitesimal version in terms of $\Gamma$ and 
$\Gamma_{x_{0}}$. More generally, for any connected group action with 
connected isotropy whose infinitesimal generators are \(gamma3323) and 
\(iso3323), 
respectively, our PSC results remain valid.

The general form of a $\Gamma$-invariant metric is given by 
$$
\eqalign{
g(q) &= \half q^{1} \left(dx^{1}\otimes dx^{1}+dx^{3}\otimes dx^{3}
\right) + q^{2}(dx^{2} +  x^{1} dx^{3} )\otimes (dx^{2} +  x^{1} 
dx^{3} )\cr
&\quad+ q^{3} (dx^{2}\odot dx^{4} +  x^{1}dx^{3}\odot dx^{4})
+ q^{4} dx^{4}\otimes dx^{4}\cr 
&= q^{i} h_{i},}
$$
where $q^{i}$ are constants subject to
$$
(q^1)^2(4\,q^{2}q^{4}-(q^3)^2) <0.
$$

It is easily verified that the infinitesimal linear isotropy representation is, 
at 
each point, that of the Lorentz subalgebra of rotations.  
This guarantees that the infinitesimal form of the Palais 
condition is satisfied so that PSC2 is valid. The chain
$$
\chi = \partial_{1}\wedge \partial_{2}\wedge 
\partial_{3}\wedge\partial_{4}.
\tag chi3223
$$
is $\Gamma$-invariant and satisfies $L_{\partial\over\partial x^{2}}\chi=0$ 
so that, according to Proposition
5.4, a cochain map exists and PSC1 is valid. Thus PSC is valid for this 
group action.

The reduced Einstein tensor is of the form
$$
\G(g(q)) = \G_{i}(q) f^{i},
$$
where
$$
f^{i} = \left(\partial_{1}\otimes\partial_{1} 
+(x^{1})^{2} \partial_{2}\otimes\partial_{2} - 2x^{1} 
\partial_{2}\odot\partial_{3}
+\partial_{3}\otimes\partial_{3},
\partial_{2}\otimes\partial_{2},
2\partial_{2}\odot\partial_{4},
\partial_{4}\otimes\partial_{4}\right),
\tag f3323
$$
and
$$
\G_{1} = -2{q^{2}\over (q^{1})^{3}},
$$
$$
\G_{2} = -2 {6 q^{2} q^{4} - (q^{3})^{2}\over (q^{1})^{2}((q^{3})^{2}- 
4 q^{2} q^{4}) },
$$

$$
\G_{3} = 2{q^{2}q^{3}\over (q^{1})^{2}  ((q^{3})^{2}- 
4 q^{2} q^{4})},
$$
$$
\G_{4} = -4 {(q^{2})^{2}\over (q^{1})^{2}  ((q^{3})^{2}- 
4 q^{2} q^{4})}.
$$

We note that there are no $G$-invariant solutions to the vacuum 
Einstein equations $\G=0$ in this example. 
One can get equations that admit solutions 
by adding a $G$-invariant energy momentum tensor, which we won't 
bother to do. In any case, our goal here is to 
 verify the equivalence of the reduced field equations with 
 the Euler-Lagrange expressions coming from the reduced 
Lagrangian, which is guaranteed by the absence of the two obstructions to 
PSC. If we were viewing PSC as a statement about 
$G$-invariant critical points of an action functional, 
then  we would have to admit that, in this case, there 
are no $G$-invariant critical points, which could be the case whether or not 
the reduced equations were in fact equivalent in the sense of
Appendix B. This highlights a key 
difference between our definition of PSC and that of Palais in 
\refto{Palais1979}.

Using the cochain \(chi3223), the reduced Lagrangian is an ordinary function 
of the $q^{i}$ given by
$$
\hat\lambda(q)
=-\half\sqrt{((q^3)^2-4q^2q^4)}{q^2\over q^1}.
$$
It is easily checked that the four Euler-Lagrange expressions 
associated with $\hat \lambda$ yield the 4 components of the Einstein 
tensor. In particular, we have that
$$
E_i(\hat \lambda) ={\partial 
\hat\lambda\over\partial q^{i}}= -\sqrt{-g(q)}\,\G_{i}(g(q)).
$$

\subhead{Example 6.6b}

Finally we consider a transitive group action for which both PSC1 and PSC2 fail.  
The group $G$ with local coordinates $\lambda^{\mu}, \mu =1,\ldots,5 $,
 $\lambda ^{\mu} \in \reals$, acts on $\reals^{4}$ with coordinates $x^{\alpha}$ by
$$
\mu(\lambda,x) = ( e^{l \lambda^5}x^1 -\frac{1}{2} e^{\epsilon \lambda^5} 
(\lambda^4)^2 x^2-e^{k \lambda^5} \lambda^4 x^3 - \lambda^3, 
e^{\epsilon \lambda^5} x^2 + \lambda^1, 
 \lambda^2+e^{\epsilon \lambda^5} \lambda^1 x^2 + e^{k \lambda^5} x^3 , 
x^4 +\lambda^5 )
$$
where $k$, $l$ and $\epsilon$ are parameters satisfying $2k=l+\epsilon$.
A basis for the Lie algebra of infinitesimal generators for this action is
given by (33.29) in \refto{Petrov1969}.

Let $x^\alpha_0 = (x^1_0,x^2_0,x^3_0,x^4_0)$ be a point in $\reals^4$, then 
the action of $G_{x_0} =\reals $ is given by
$$
\mu^\alpha(\lambda,x) = ( x^1 -\frac{1}{2} (\lambda^4)^2 (x^2-x^2_0)- 
\lambda^4 (x^3-x^3_0),  x^2 , 
x^3+  \lambda^4 (x^2-x^2_0)  , x^4 )
$$
where $\lambda^4$ is a coordinate for $ G_{{\bf x}_0}$. 
The linear isotropy representation at the point $x_{0}^{\alpha}$ is
given by the matrix
$$
\left(\matrix{1&-x^2_0(\lambda^4)^2&-\lambda^4&0\cr
0&1&0&0\cr
0&\lambda^4&1&0\cr
0&0&0&1}\right).
\tag{isorep7}
$$

Using the isotropy representation \(isorep7) in the constraint \(iso_cond), 
the most general $G$-invariant metric is found to be
$$
\eqalign{
g &= 
q^1\big(e^{-2kx^{4}} dx^{1}\odot dx^{2} + \half e^{-2kx^{4}} 
dx^{3}\otimes dx^{3}\big)
+
q^2e^{-2\epsilon x^{4}} dx^{2}\otimes dx^{2}\cr
&\quad+ q^3
e^{-\epsilon x^{4}}dx^{2}\odot dx^{4}
+
q^4 dx^{4}\otimes dx^{4}\cr
&=q^{i} h_{i},}
\tag Sigma3329
$$
where the $h_{i}, i=1,\ldots,4$ are a basis for the $G$-invariant symmetric 
$\left({}^0_2\right)$ tensor fields.

It is easy to see from \(isorep7) that at each point $x\in M$ the vector 
$(\partial_1)_x \in (T_xM)^{G_x}$, and then from \(Sigma3329) that this 
vector is null with  respect to all the invariant metrics. Therefore, by 
Proposition 5.6, the Palais condition is not satisfied for this group 
action and PSC2 fails.

Let us now consider PSC1. We shall see that PSC1 is valid if and only if  $k=0$. 
It is easy to show that all $G$-invariant chains are proportional to
$$
\chi = e^{3kx^{4}}\partial_{1}\wedge
\partial_{2}\wedge
\partial_{3}\wedge
\partial_{4}.
\tag chi3329
$$
The vector space of $G$-invariant vector fields is two-dimensional and 
is spanned by
$$
S_{1} = e^{lx^{4}}\partial_{1},\quad S_{2}=\partial_{4}.
$$
Since
$$
L_{S_{1}}\chi = 0,\quad L_{S_{2}}\chi = 3k\chi,
$$
we see that in accordance with Proposition 5.4 a cochain map exists 
if and only if $k=0$.

Let us consider the reduced Einstein equations with cosmological 
constant $\Lambda$ 
for the $G$-invariant spacetimes. We have that
$$
\G(g(q)) + \Lambda g^{-1}(q)  = \Delta_{i}(q)f^{i},
$$
where $f^{i}$ are a basis for the $G$-invariant symmetric 
$\left({}^2_0\right)$ tensor fields
$$
f^i =  \big({4\over 3} e^{2kx^{4}}\partial_{1}\odot \partial_{2}
+{2\over 3}e^{2kx^{4}}\partial_{3}\otimes\partial_{3},
e^{2lx^{4}}\partial_{1}\otimes\partial_{1},2e^{lx^{4}}\partial_{1}
odot\partial_{4},
\partial_{4}\otimes \partial_{4}\big). 
$$
and
$$
\eqalign{
\Delta_{1}&={9 k^{2} + 3 \Lambda q^{4}\over q^{1}q^{4}}\cr
\Delta_{2}&=- {2 q^{2}q^{4}(l^{2} +l \epsilon + 4\epsilon^{2}) - 
3(q^{3})^{2}k^{2}
+ \Lambda q^{4}( 4q^{2}q^{4} -  (q^{3})^{2})
\over 4 (q^{1}q^{4})^{2}}\cr
\Delta_{3}&=-{q^{3}(3k^{2}+ \Lambda q^{4})\over q^{1}(q^{4})^{2}}
\cr
\Delta_{4}&={3k^{2}+ \Lambda q^{4}\over (q^{4})^{2}}.
}
$$

Using the $G$-invariant chain $\chi$ in \(chi3329) to construct the reduced 
Lagrangian yields
$$
\hat\lambda(q)=-\sqrt{2 (q^{1})^{3}\over q^{4}}(3k^{2} + \half \Lambda 
q^{4}).
$$
Since $\hat\lambda$ is independent of $q^{2}$ and $q^{3}$ it 
is clear that two of the Euler-Lagrange 
expressions (now just partial derivatives) will be trivial:
$$
E_2(\hat\lambda) = {\partial \hat 
\lambda\over \partial 
q^{2}} = 0,\quad E_3(\hat\lambda) = {\partial \hat 
\lambda\over \partial 
q^{3}} = 0.
$$
The corresponding field equation components are $\Delta_{2}$ and 
$\Delta_{3}$, both  of which are non-zero. This is how 
the failure of PSC2 manifests 
itself. The remaining Euler-Lagrange expressions are
$$
E_1(\hat\lambda) = {\partial \hat 
\lambda\over \partial 
q^{1}} = -3{\sqrt{1 \over 2(q^{1})^{3}q^{4}}}(3k^{2} + \half \Lambda 
q^{4}) 
\tag el33291
$$
and
$$
E_4(\hat\lambda) = {\partial \hat 
\lambda\over \partial 
q^{4}} = -{\sqrt{1 \over 2q^{1}(q^{4})^{3}}}(-3k^{2} + \half \Lambda 
q^{4}) .
\tag el33292
$$
By comparing \(el33291) with $\Delta_{1}$ and \(el33292) with 
$\Delta_{4}$  it is clear that these Euler-Lagrange expression of $\hat 
\lambda$ agree with the corresponding terms in the field equations if 
and only if $k=0$, which is the case where PSC1 holds.

\bigskip\bigskip
\noindent{\bf Acknowledgments}

We would like to thank Ian Anderson for essential contributions to many 
aspects of the problems considered here. This work was supported in 
part by NSF grants DMS-9804833, 
PHY-9732636 and PHY-0070867 to Utah State 
University. 

\bigskip\bigskip

\taghead{A.}
\head{Appendix A: Additional fields beyond the metric}

Much of the preceding analysis can be generalized to a gravitational 
field theory involving a metric coupled to matter fields, or to a 
matter field theory propagating on a fixed spacetime. We 
briefly indicate here how this generalization goes (see also \refto{AFT2000}).

Let us collectively denote all dynamical fields on a spacetime manifold 
$M$ using an abstract 
symbol $\varphi$. We assume that all fields can be viewed as sections of fiber 
bundles $E\to M$. If the theory is that of
a  gravitational field coupled to matter then $\varphi$ would include the 
metric, scalar fields, Maxwell fields, \etc If the theory is a field 
theory on a fixed background spacetime $(M,g)$, then $\varphi$ 
includes just the dynamical fields. 
We assume that there is given a group action of $G$ on $M$ with a specified lift to the space of 
fields $\varphi$ obeying the regularity conditions of \refto{CGT2000},
and that there exist $G$-invariant fields $\varphi(\q)$ with 
respect to that group action. Here we denote by $\q$ the fields on 
$M/G$ that parametrize the space of $G$-invariant fields on $M$. For 
more details about all these constructions, see \refto{CGT2000}.

Let $\lambda$ be a $G$-equivariant Lagrangian for the fields $\varphi$. 
The first variational 
formula
$$
\delta {\lambda} =  E(\lambda)\cdot \delta\varphi
+ d \eta(\delta\varphi),
$$
defines $\eta$ up to an exact form. We assume that the $(n-1)$-form $\eta$ can be 
chosen to be $G$-equivariant. For a generally covariant theory 
of gravity coupled to matter fields this is guaranteed \refto{Iyer1994}. If the 
metric is fixed, \ie we consider matter fields on a fixed spacetime, 
then this will again be guaranteed provided chosen symmetry group acts by
 isometries 
\refto{AFT}. We would normally need this latter assumption in any 
case in order to render the matter Lagrangian $G$-equivariant. 

With 
this assumption in hand, we can repeat the analysis of sections 4 and 
5, with very similar results. As in \S 4 we can define 
field equations, reduced field equations and the reduced Lagrangian 
associated with the $G$ action. As in \S 5 
we can obtain the two conditions (PSC1 and PSC2) for PSC. 
The condition PSC1 is described in terms of the Lie 
algebra cohomology exactly as before: ${\cal H}^{l}(\Gamma,G_{x})\neq 0$ 
for all $x\in M$,
where $l$ is the dimension of the group orbits in $M$. 
Thus PSC1 is insensitive to the field content of the theory. 
On the other hand, the validity of PSC2 
depends upon the way in which the isotropy group $G_{x}$ is 
represented  on $S_x$,  the vector space of values  of the 
field variations $\delta\varphi$ at 
each $x\in M$.  As in \S 5, denote by $\sgx$ the $G_{x}$-invariant
points in $S_x$ and denote by $\sgstarx$ the 
$G_{x}$-invariant points in $S_x^{*}$. 
Denote by $\sgxo\subset S_x^{*}$ the annihilator of 
$\sgx$. As before,  PSC2 is satisfied if and only 
if
$$
\sgstarx\cap \sgxo = 0,\quad \forall\ x\in M.
\tag generalintersect
$$

A couple of examples of \(generalintersect) are worth mentioning. For 
a theory involving scalar fields coupled to gravity, so that $\varphi$ includes a 
metric and some scalar fields, the action of $G_{x}$ on the 
scalar fields is trivial and so the condition \(generalintersect) 
reduces to that studied in \S 5 on the metric alone. So, for example, 
in four dimensions with  connected isotropy groups only the null 
rotation subgroups lead to a failure of PSC2  for a metric 
coupled to  scalar fields. For scalar  fields propagating on a fixed 
($G$-invariant)
background spacetime the action of $G_{x}$ on $S_x$ is completely trivial so 
that 
$$
\sgx =S_x,
$$
and it follows that \(generalintersect) is always satisfied. As 
another example, suppose that $\varphi$ includes a Maxwell field and a 
metric, or just a Maxwell field propagating on a fixed spacetime. 
For simplicity, let us assume that the $G$-action on the Maxwell 
field does not include any gauge transformations, that is, a 
$G$-invariant Maxwell field is just a $G$-invariant 1-form on 
spacetime. It is not hard to see that the condition \(generalintersect) 
leads to the null rotation subgroups as the only isotropy groups that 
fail to satisfy PSC2.

\taghead{B.}
\head{Appendix B: Equivalence of differential equations}

For our purposes we will define 
``equivalent'' differential equations as follows. Let $\Delta_{1}$ and 
$\Delta_2$ be two differential operators acting on a given space of 
fields ${\cal M}$. We view them as maps :
$$
\Delta_{1}\colon {\cal M} \to V_{1},\quad \Delta_{2}\colon {\cal M} \to 
V_{2},
$$
where $V_{1}$ 
and $V_{2}$ are vector spaces.
We say that the differential equations $\Delta_{1}=0$ and 
$\Delta_{2}=0$ are {\it equivalent} if there is a smooth map
$$
H\colon {\cal M}\times V_{1}\to V_{2},
$$
locally constructed from the fields $\varphi\in {\cal M}$, such that 
for each $\varphi$
$$
\Delta_{2}(\varphi) = H(\varphi) \cdot \Delta_{1}(\varphi),
\tag equiv
$$
with $H(\varphi)\colon V_{1}\to V_{2}$ an isomorphism. In this case we 
write
$$
\Delta_{1}=0\Longleftrightarrow \Delta_{2}=0.
$$
The equivalence of equations shown in 
\(reducedeq) is of the type just described.

The reduced 
differential operator $\hat \Delta$, described in \S 3, can be viewed as a mapping
$$
\hat \Delta\colon \hat{\cal Q} \to V,
$$
where $V$ is a vector space. Likewise,
$$
E(\hat\lambda)\colon \hat{\cal Q} \to W,
$$
where elements of $W$ differ from those of $V$ by a tensor product 
with a volume form on 
$M/G$. The  vector spaces $V$ and $W$ are isomorphic. PSC asserts 
that
$$
E(\hat\lambda)=0 \Longleftrightarrow \hat\Delta=0.
$$

\references
\refis{Olver1993}{P. Olver, {\it Applications of Lie Groups 
to Differential Equations}, (Springer-Verlag, New York 
1993).}

\refis{Pauli1921}{W. Pauli, {\it Relativit\"atstheorie}, Encyclopedia der 
Matematicshcen Wissenschaftern, Vol. 19, B.G. Teubner, Leipzig, 1921.}

\refis{MacCallum1971}{M. MacCallum, \cmp 20, 57, 1971.}
 
\refis{Ryan1974}{M. Ryan, \journal J. Math. Phys., 15, 812, 
1974.}

\refis{MacCallum1979}{M. MacCallum in {\it 
General 
Relativity: An 
Einstein Centenary Survey}, edited by S. Hawking and W. 
Israel 
(Cambridge University Press, Cambridge 1979).}

\refis{Lovelock1973}{D. Lovelock, \journal Nuov. Cim., B73, 
260, 1973.}

\refis{Hawking1969}{S. Hawking, 
\journal Mon. Not. R. Astron. Soc., 142, 129, 1969.}

\refis{CGT1996}{J. D. Romano and C. G. Torre, 
\prd 53, 5634, 1996.}

\refis{Ashtekar1991c}{A. Ashtekar and J. Samuel, \cqg 8, 2191, 1991.}

\refis{KN1969}{S. Kobayashi and K. Nomizu, {\it Foundations of 
Differential Geometry}, Vol 2, (Wiley, 1969).}

\refis{Shepley1998}{J. Pons and L. Shepley,  \prd 
58, 024001, 1998.}

\refis{Anderson1997}{I. M. Anderson and M. E. Fels,
\journal Amer. J. Math., 119, 609, 1997.}

\refis{Sneddon1976}{G. Sneddon, \journal J. Phys. A, 9, 229, 
1976.}

\refis{MacCallum1972}{M. MacCallum and A. Taub, \journal 
Commun. Math. Phys., 25, 173, 1972.}

\refis{Palais1979}{R. Palais, \cmp 69, 19-30, 1979.}

\refis{CGT2000}{I. M. Anderson, M. E. Fels, and C. G. Torre, \cmp 212, 
653-686, 2000.}

\refis{Petrov1969}{A. Petrov, {\it Einstein Spaces}, 
(Pergamon Press, 1969).}

\refis{Coquereaux1988}{R. Coquereaux and A. Jadczyk, {\it Riemannian 
Geometry, Fiber Bundles, Kaluza-Klein Theories and all that}, Lecture 
Notes in Physics {\bf 16} (World Scientific, Singapore, 1988).}

\refis{Anderson2001}{I. M. Anderson and M. E. Fels, ``Transverse 
group actions on bundles'', to appear in {\sl Topology and its 
Applications}, 2001.}

\refis{Iyer1994}{V. Iyer and R. Wald, \prd 50, 846-864, 1994.}

\refis{Winternitz1975}{J. Patera and P. Winternitz, \jmp 16, 1597, 1975.}

\refis{Helgason2001}{S. Helgason, {\it Differential Geometry, Lie 
Groups, and Symmetric Spaces}, (American Mathematical Society, 2001).}

\refis{MacCallum1980}{D. Kramer, H. Stephani, E. Herlt, M. MacCallum, 
{\it Exact Solutions of Einstein's Field Equations}, edited by E. 
Schmutzer (Cambridge University Press, Cambridge, 1980).}

\refis{CGT1996}{J. D. Romano and C. G. Torre, 
\prd 53, 5634, 1996.}

\refis{AFT}{I. M. Anderson, M. E. Fels, C. G. Torre, unpublished.}

\refis{AFT2000}{I. M. Anderson, M. E. Fels, C. G. Torre,
``Group Invariant Solutions without 
Transversality and the Principle of Symmetric Criticality'', preprint 
math-ph/9910014.}

\refis{Jantzen1980}{R. T. Jantzen, \journal {\sl Nuovo Cim.}, 
55B, 161-172, 1980;
\prd 34, 424-433, 1986;
\prd 35, 2034-2035, 1987.}

\endreferences

\bye